\documentclass[amsmath,amssymb,aps,showpacs,onecolumn,superscriptaddress,letterpaper,longbibliography, floatfix]{revtex4-2}
\usepackage[document]{ragged2e}
\usepackage[utf8]{inputenc}
\usepackage{amsmath}
\usepackage{amssymb}
\usepackage{graphicx}
\usepackage{xcolor}
\usepackage{hyperref}
\usepackage{ulem}
\usepackage{xspace}
\usepackage{siunitx}
\usepackage{color}
\usepackage{comment}

\usepackage{mciteplus}

\mciteErrorOnUnknownfalse

\newcommand{\nubar}[0]{\overline{\nu}}
\newcommand{\cevns}{CE$\nu$NS\xspace}

\usepackage[textsize=tiny,%
    color=orange!25,%
    linecolor=orange,%
    bordercolor=lightgray]{todonotes}
\usepackage{tabularx} 
\usepackage{booktabs} 
\usepackage{lineno}

\usepackage{titlesec}
\titlespacing*{\section}{0pt}{0.9\baselineskip}{0.9\baselineskip}

\newlength{\parenbarKernelHeight}
\providecommand{\parenbar}[1]{%
    \settoheight{\parenbarKernelHeight}{\ensuremath{#1}}%
    \addtolength{\parenbarKernelHeight}{0pt}
    \llap{\raisebox{\parenbarKernelHeight}{\scalebox{0.5}[0.5]{(}}}%
    \overline{#1}%
    \rlap{\raisebox{\parenbarKernelHeight}{\scalebox{0.5}[0.5]{)}}}%
}

\providecommand{\nuany}{\ensuremath{\hspace{0.5mm}\parenbar{\nu}}\xspace}
\providecommand{\numuany}{\ensuremath{\nuany_\mu}\xspace}
\providecommand{\nueany}{\ensuremath{\nuany_{e}}\xspace}

\begin{document}

\title{Neutrino Scattering: Connections Across Theory and Experiment}

\newcommand{\Kentucky}{Department of Physics and Astronomy, University of Kentucky, Lexington, KY 40506, USA}
\newcommand{\LANL}{Los Alamos National Accelerator Laboratory, Los Alamos, NM, USA}
\newcommand{\FNAL}{Fermi National Accelerator Laboratory, Batavia, IL, USA}
\newcommand{\IFIC}{Instituto de F\'{i}sica Corpuscular, CSIC and Universidad de Valencia, E-46980, Spain}
\newcommand{\Ghent}{Ghent University, Department of Physics and Astronomy, B-9000 Ghent, Belgium}
\newcommand{\Virginia}{Center for Neutrino Physics, Virginia Tech, Blacksburg, Virginia 24061, USA}
\newcommand{\Granada}{Departamento de F\'isica
At\'omica, Molecular y Nuclear, Universidad de Granada, E-18071, Spain}
\newcommand{\CERN}{CERN, European Organization for Nuclear Research, Geneva, Switzerland}
\newcommand{\Oxford}{University of Oxford, Oxford, OX1 3RH, United Kingdom}
\newcommand{\Torino}{Dipartimento di Fisica, University of Turin, and INFN, Turin, I-10125 Italy}
\newcommand{\IPSA}{IPSA-DRII, 94200 Ivry-sur-Seine, France}
\newcommand{\Sorbonne}{Sorbonne Universit\'e, Universit\'e Paris Diderot, CNRS/IN2P3, Laboratoire
de Physique Nucl\'eaire et de Hautes Energies (LPNHE), Paris, France}
\newcommand{\Argonne}{Argonne National Laboratory, Lemont, Illinois 60439 USA}
\newcommand{\Aligarh}{Department of Physics, Aligarh Muslim University, Aligarh-202002, India}
\newcommand{\Mainz}{Institute for Nuclear Physics, Johannes Gutenberg University, 55128, Mainz, Germany}
\newcommand{\Mainza}{Institute for Physics, Johannes Gutenberg University, 55128, Mainz, Germany}
\newcommand{\Madrid}{Grupo de F\'isica Nuclear, Universidad Complutense de Madrid and IPARCOS,  Madrid, Spain}
\newcommand{\StLouis}{Washington University in St. Louis and the McDonnell Center for the Space Sciences, St. Louis, Missouri, 63130, USA}
\newcommand{\Lancaster}{Physics Department, Lancaster University, Lancaster, United Kingdom, LA1 4YB}
\newcommand{\Michigan}{Michigan State University, East Lansing, Michigan, 48824, USA}
\newcommand{\Giessen}{Institut fuer Theoretische Physik, Universitaet Giessen, D-35392 Giessen, Germany}
\newcommand{\Duluth}{Department of Physics, University of Minnesota – Duluth, Duluth, Minnesota 55812, USA}
\newcommand{\Minnesota}{Department of Physics, University of Minnesota Twin Cities, Minnesota Minnesota 55455, USA}
\newcommand{\CNRS}{CNRS, Université Paris Cité, Astroparticule et Cosmologie, F-75013 Paris, France}
\newcommand{\Warwick}{University of Warwick, Coventry, CV4 7AL, United Kingdom}
\newcommand{\London}{King's College London, London WC2R 2LS, United Kingdom}
\newcommand{\Geneva}{University of Geneva, Section de Physique, DPNC, Geneva, Switzerland}
\newcommand{\Wroclaw}{Institute of Theoretical Physics, University of Wroc\l aw, Wroc\l aw, Poland}
\newcommand{\LLNL}{Nuclear and Chemical Sciences Division, Lawrence Livermore National Laboratory, Livermore, CA 94550, USA}
\newcommand{\Technion}{Technion, Physics Dept.,  Israel Institute of Technology, Haifa, Israel}
\newcommand{\IPPP}{Institute for Particle Physics Phenomenology, Durham University, SouthRoad, DH1 3LE,Durham, UK}
\newcommand{\IFT}{Instituto de F\'isica Te\'orica UAM-CSIC, Calle Nicol\'as Cabrera 13-15, 28049 Madrid, Spain}
\newcommand{\Imperial}{Department of Physics, Imperial College London, London, UK}
\newcommand{\Sevilla}{Departamento de FAMN, Universidad de Sevilla, Spain}
\newcommand{\Pittsburgh}{PITT PACC, Department of Physics and Astronomy, University of Pittsburgh, USA}
\newcommand{\WandM}{Department of Physics, William \& Mary, 
Williamsburg, Virginia 23187, USA}
\newcommand{\CNP}{Center for Neutrino Physics, Virginia Tech, Blacksburg, VA 24061, USA}
\newcommand{\TAU}{Department of Particle Physics, Tel Aviv University, Tel Aviv-Yafo 6997801, Israel}
\newcommand{\Duke}{Department of Physics, Duke University, Durham, NC 27708, USA}
\newcommand{\INT}{Institute for Nuclear Theory, University of Washington, Seattle, WA 98195, USA}
\newcommand{\Udub}{University of Washington, Seattle, WA 98195, USA}
\newcommand{\RBI}{Rudjer Boskovic Institute, 10000 Zagreb, Croatia}
\newcommand{\Rochester}{University of Rochester, Rochester, NY 14627,USA}
\newcommand{\Sinica}{Institute of Physics, Academia Sinica, Taipei City, 11529, Taiwan}


\author{L.~Alvarez-Ruso}\affiliation{\IFIC}
\author{A.M. Ankowski}\affiliation{\Wroclaw}
\author{A.~Ashkenazi}\affiliation{\TAU}
\author{J.~Barrow}\affiliation{\Minnesota}
\author{M.~Betancourt}\affiliation{\FNAL}
\author{K.~Borah}\affiliation{\Kentucky}
\author{M.~Sajjad Athar}\affiliation{\Aligarh}
\author{E.~Catano-Mur} \affiliation{\WandM}
\author{P.~Coloma}\affiliation{\IFT}
\author{P.~Dunne}\affiliation{\Imperial}
\author{L.~Doria}\affiliation{\Mainz}
\author{A.~Fedynitch}\affiliation{\Sinica}
\author{A.~Garcia-Soto}\affiliation{\IFIC}
\author{S.~Gardiner}\affiliation{\FNAL}
\author{R.~Gonz\'alez-Jim\'enez}\affiliation{\Sevilla}
\author{P.~Huber}\affiliation{\CNP}
\author{N.~Jachowicz}\affiliation{\Ghent}
\author{E.~Kajomovitz}\affiliation{\Technion}
\author{B.~Klicek}\affiliation{\RBI}
\author{J.~Kopp}\affiliation{\CERN}\affiliation{\Mainza}
\author{K.~Long}\affiliation{\Imperial}
\author{I. Martinez-Soler}\affiliation{\IPPP}
\author{A.~S.~Meyer}\affiliation{\LLNL}
\author{C.~Marshall}\affiliation{\Rochester}
\author{L.~Munteanu}\affiliation{\CERN}
\author{A.~Nikolakopoulos}\affiliation{\Udub}
\author{V.~Pandey}\affiliation{\FNAL}
\author{A.~Papadopoulou}\affiliation{\LANL}
\author{M.~Scott}\affiliation{\Imperial}
\author{N.~Rocco}\affiliation{\FNAL}
\author{K. Scholberg}\affiliation{\Duke}
\author{J.E.~Sobczyk}\affiliation{\Mainz}
\author{C.~Wret}\affiliation{\Imperial}
\author{Z.~Tabrizi}\affiliation{\Pittsburgh}\affiliation{\CERN}

\begin{abstract}
\justify
One of the highest priorities in fundamental particle physics is the precise measurement of neutrino properties and interactions. The discovery of non-zero neutrino masses remains one of the very few hints regarding the nature of physics beyond the Standard Model (SM).
Fundamental questions about neutrino properties still remain unanswered, such as their absolute masses, the ordering of the mass states and the charge-parity (CP) violating phase.
Moreover, neutrinos have the potential to carry information regarding the nature of astrophysical sources. They are also a tool to  expand our understanding of nuclear structure. 
To investigate these questions, a large number of ongoing and future experiments probe neutrino interactions with matter in a broad range of energies scales and a variety of targets. 
They are sensitive to different neutrino-induced reaction mechanisms, challenging nuclear theory at various energy regimes and require a good understanding of electroweak response of nuclei from low-lying excitations, through the quasielastic regime and the baryon resonance excitation region up to deep inelastic scattering.

{\bf In this document, put forward by the Neutrino Scattering Theory Experiment Collaboration (NuSTEC), we provide input on the experimental and theoretical efforts towards accurate and precise understanding of neutrino interactions, which is a crucial requirement for a number of programs across many energy scales.} The neutrino oscillation program is one of the driving forces of these efforts. Among accelerator-based neutrino experiments, the Deep Underground Neutrino Experiment (DUNE)~\cite{dune_1,dune_2,dune_3,dune_4} and Hyper-Kamiokande (HK)~\cite{ESPP_hk} will propel the field into the precision era in the near future. This is aligned with 2024 Strategic Plan for U.S. Particle Physics (P5), as well as with the strategy of particle physics community in Japan which has identified the Tokai-to-Hyper-Kamiokande project as one of its highest particle physics priorities for the next decade.
The European experimental and theoretical engagement in various aspects of these programs is invaluable, as explained in more detail in the document. 
In this context, it is important to highlight a European proposal for the ESSnuSB next-to-next generation neutrino oscillation experiment~\cite{ESPP_ESSnuSB}.
%

%

Long-baseline oscillation experiments also conduct an extensive program focused on studying neutrino-nucleus cross sections using their near detectors. 
In similar energy ranges, the SBN program at CERN~\cite{ESSP_SBN} and at the Fermi National Accelerator Laboratory (FNAL)~\cite{ESSP_SBN_FNAL}, as well as the upgraded T2K experiment~\cite{ESPP_t2k}, will provide additional insights.
The CEvNS program~\cite{ESSP_coherent} and ESSnuSB at lower energies, as well as KM3NeT and IceCube at higher energies, further contribute to our understanding of neutrino-nucleus interactions.

Furthermore, there are ongoing European initiatives that aim to deliver well-understood neutrino beams that would allow precise measurements of cross sections from NuSTORM~\cite{ESSP_nustorm} and FASER$\nu$. 
At the same time, European facilities deliver precise electron and hadron-induced reaction data from ProtoDUNE, as well as dedicated electron scattering exepriments to improve neutrino energy reconstruction techniques through external data validation.


Future breakthroughs in the field critically require improvements in the theoretical modeling of neutrino-nucleus cross sections. 
In Europe, diverse groups are developing highly realistic models tailored to different energy ranges and relevant degrees of freedom.
Yet, more work is needed to support experiments, especially in the high-energy sector and in uncertainty quantification. 
In parallel, Monte Carlo event generators are evolving into flexible interfaces that incorporate advanced theoretical models in order to translate raw detector signals into measurable cross sections. 
{\bf Next-generation precise measurements encourage strengthening the connection between Beyond Standard Model searches and neutrino cross section predictions~\cite{ESSP_nu_theory}}.
It should be noted that many potential Beyond Standard Model effects in neutrino oscillations (neutrino decay, non-standard interactions, non-unitarity of the PMNS matrix) are degenerate with standard oscillations.  
Detecting these sub-leading effects requires data from multiple experiments, which in turn demands substantial advancements in theoretical models and event generators to achieve the necessary consistency and precision.


 

This broad field of experimental and theoretical activities is supported by many European institutions and experts.
These include particle beams, water Cherenkov and/or the liquid argon time projection chamber (LArTPC) technology; facilities performing electron and hadron scattering; and a strong neutrino and nuclear theory community developing models of neutrino interaction with nuclei and hadrons. 
{\bf Precision-era neutrino research demands robust collaboration between theory and experiment, yet current efforts remain fragmented with no clear leadership or sufficient institutional support.} 
Establishing a CERN Neutrino Physics Center~\cite{ESSP_nu_theory,ESSP_nu_CERN}—modeled on FNAL’s successful LHC Physics Center—would create a unified European hub for both experimental and theoretical neutrino activities. This center, evolved from the Neutrino Platform at CERN, is envisioned to drive coordinated scientific leadership, enhance prioritization, and provide the critical precision theory predictions and simulation support necessary to advance neutrino physics.

\end{abstract}


\maketitle


\justify
\section{Introduction}

The precise measurement of neutrino properties is among the highest priorities in fundamental particle physics, involving many experiments worldwide. 
Since the experiments rely on the interactions of neutrinos with bound nucleons inside atomic nuclei, the planned advances in the scope and precision of these experiments require a commensurate effort in the understanding and modeling of the hadronic and nuclear physics of these interactions, which is incorporated as a nuclear model 
in neutrino event generators.  This model is essential to every phase of experimental analyses and its theoretical uncertainties play an important role in interpreting every result.
Here we discuss  the impact of neutrino–nucleus interactions
on the measurement of neutrino properties using the determination of oscillation parameters as a central example. 
This discussion includes an overview of the experimental challenges, how event generators work, and the status of the relevant theory.
We then emphasize how our understanding must improve to meet the demands of future experiments. In every topic we find that the challenges can be overcome only through the active support of integrated collaboration across strong and electroweak physics that includes theorists and experimentalists from both the nuclear and high energy physics communities.
Further details on the experiments, event generators, and relevant theory precision can be found in reviews, white papers, and references therein.



\section{Experimental efforts}

The experimental study of neutrino oscillations spans a wide range of experiments, each designed to constrain different oscillation parameters by examining the appearance or disappearance of specific neutrino flavors across different energy regions. 
All relevant experiments rely on heavy nuclei as targets and, thus, their accuracy is limited by our understanding of neutrino-nucleus interactions at various energy scales.

There is a number of key areas which need to be addressed in preparation for the next-generation experiments' program, one of particular importance is the relationship between the true and reconstructed neutrino energy. Nuclear effects alter the amount of visible energy in a detector and can impact the shape of the oscillated spectrum in a way which is degenerate with neutrino oscillations.
It is inevitable that a fraction of the neutrino energy will be invisible in the detector 
so the uncertainty introduced by the difference between true and reconstructed neutrino energies  must be properly quantified and reduced to a level compatible with the target precision on oscillation measurements.
Furthermore, the differences between the cross sections for different neutrino flavors and neutrinos and anti-neutrinos need to be understood.
In principle, these differences are predictable thanks to the assumption of lepton universality. However, mass differences between lepton flavors become significant in areas of the interaction phase space which have large associated uncertainties.
A typical example would be at low energy transfers, where nuclear models differ substantially in the way in which they incorporate the treatment of final state interactions, optical potential, Pauli blocking or long-range correlations~\cite{Nikolakopoulos:2019qcr}.
Finally, it is important to study the energy dependence of the cross section. Due to oscillations, the flux probed by near detectors is different from the one seen at the far detector. 
To reliably extrapolate near detector constraints to the far detector, the evolution of the cross section must be known for the range of energies relevant for oscillation searches. Therefore, it is essential to have high-quality neutrino scattering measurements in order to benchmark the nuclear interaction model. 

This section offers an overview of various experiments, outlining their objectives, requirements, and the complementary measurements they can provide. 
The experiments are arranged in order of increasing incoming neutrino energy. 
To address some of the challenges in neutrino interaction modeling, the final subsection highlights dedicated experiments that aim to refine these models and reduce uncertainties using both neutrino and electron scattering data by leveraging nuSTORM and dedicated electron scattering experiments.

\textbf{CE$\nu$NS:} A number of experiments worldwide are performing Coherent Elastic Neutrino-Nucleus Scattering (CE$\nu$NS) measurements using spallation sources, astrophysical neutrinos, and reactors.
Neutrinos from pion decay at rest provide a well-understood source, in both spectrum and time with respect to a pulsed beam, of neutrinos of three flavors.
These include monochromatic $\nu_\mu$ from pion decay, and $\bar{\nu}_\mu$, $\nu_e$ from three-body muon decay.
Their energies are in the range of tens of MeV up to about 50~MeV.  
Since the maximum CE$\nu$NS recoil energy scales as $\sim \sqrt{2 E_\nu^2/M}$, where $M$ is the mass of the recoiling nucleus, stopped-pion neutrinos result in tens of keV recoils for medium-size nuclei.  
The COHERENT experiment has exploited these neutrinos at the Spallation Neutron Source's pulsed beam at OakRidge National Laboratory for CE$\nu$NS measurements on three materials to date, namely CsI~\cite{PhysRevLett.129.081801}, Ar~\cite{PhysRevLett.126.012002}, and Ge~\cite{adamski2024detectioncoherentelasticneutrinonucleus}.  
More measurements are ongoing, with the aim to map out the target-mass dependence of the cross section.  
In Europe, there are plans to use the European Spallation Neutron Source for CE$\nu$NS studies, using several detectors.

At slightly lower neutrino energy, dark matter detectors are designed to discover  Weakly Interacting Massive Particles (WIMPs) with masses in the GeV regime.
These are also sensitive to competing recoils from other astrophysical sources such as solar neutrinos. 
Indeed, two xenon-based experiments~\cite{PhysRevLett.126.091301,PhysRevLett.133.191002} have observed CE$\nu$NS from solar neutrinos, providing the first hints of the so-called ``neutrino fog". 
At even lower neutrino energies of several MeV, recoil energies are in the sub-keV range and experimentally extremely challenging to detect.  
During the past decades, a very wide range of detector
technologies has been deployed in search of reactor-neutrino CE$\nu$NS,
culminating in a successful first measurement by CONUS+~\cite{Ackermann:2025obx}, which makes
use of high-purity Ge diode detectors.  
Many other searches are underway at many reactor sites.
Radioactive sources have also been proposed for ultra-low-energy-recoil CE$\nu$NS searches, but come with great technical challenges.

The higher the momentum transfer, the larger the effect of the nuclear
form factor on the recoil spectrum.  At stopped-pion neutrinos
energies, there are reasonable prospects to determine neutron form
factors; at percent-level precision one needs to disentangle nuclear
effects from potential BSM signatures in the recoil energy
distributions.  At reactor energies, the form factor is effectively
unity, providing a cleaner BSM search territory, but limiting prospects
for shedding light on nuclear structure.



\textbf{ESSnuSB:} The European Spallation Source Neutrino Super Beam (ESSnuSB)~\cite{Alekou:2022emd} is a proposed next-to-next generation neutrino oscillation experiment to precisely measure the CP-violating phase at the second oscillation maximum. 
It will use the \SI{5}{MW} European Spallation Source (ESS) proton linac \cite{Garoby:2017vew} for the neutrino beam production with the on-axis \SI{500}{kt} water Cherenkov far detectors located at a mine \SI{360}{km} from the source. The neutrino energies will be in the 60-600 MeV range. 
As the knowledge of the $\nu_\mu$ and $\nu_e$ interaction cross section with water in this energy rang will be crucial for the precision measurement, a staged approach is proposed in which a cross-section measurement will take place during the far detector construction \cite{ESSnuSB:2023ogw}. 
For that purpose, a low energy monitored beam (LEMNB) facility and a low energy neutrino beam from a ring of stored muons (LEnuSTORM) are foreseen to be built. 
The LEMNB will operate first, after which LEnuSTORM will take over. Neutrino interactions from both of these sources will be detected using a dedicated water Cherenkov detector. The LEMNB technique allows for a very precise neutrino flux measurement, hence the expected error on the $\nu_\mu$ cross-section is of the order on 1\%; the precision of the $\nu_e$ measurement is statistically limited. The LEnuSTORM facility will provide both $\nu_e$ and $\nu_\mu$ fluxes in a roughly equal part, which makes it suitable for $\nu_e$ cross-section measurement; the momentum of the stored muons will be tuned so that neutrino energies match those of the main ESSnuSB beam. Should the experiment be approved, the construction is expected to begin in the 2030s.

{\bf SBN:} At Fermi National Accelerator Laboratory (FNAL), the Short-Baseline Neutrino program (SBN) delivers a rich and compelling physics opportunity, including the ability to resolve a class of experimental anomalies in neutrino physics and to perform the most sensitive search to date for sterile neutrinos at the eV mass-scale through both appearance and disappearance oscillation channels.
The program consists of Short-Baseline  Near Detector (SBND), the MicroBooNE detector and and short base-line far detector (ICARUS), all utilising the Liquid Argon Time Projection Chamber (LArTPC) technology. 

SBND is a 112-ton LArTPC located only 110 meters away from the Booster Neutrino Beam (BNB) target~\cite{MicroBooNE:2015bmn}. 
Due to its proximity to the neutrino source, SBND will collect an unprecedentedly high event rate of neutrino interactions, providing an ideal avenue for precision studies of neutrino-argon interactions. 
SBND will record approximately $\sim$7,000 neutrino events per day, amounting to $\sim$2,000,000 $\nu_\mu$ events per year and $\sim$15,000 $\nu_e$ events per year, enabling both inclusive and exclusive neutrino-argon interaction measurements. 
Before the DUNE era, SBND will produce the world's highest-statistics neutrino-argon cross section measurements. 
Leveraging the high resolution of LArTPC technology, SBND will perform precise measurements of multiple final-state interactions for both $\nu_\mu$ and $\nu_e$ events. 
The high event rate will also enable the detection of several thousands of rare interactions, such as those associated with the production of $\Lambda$ and $\Sigma^{+}$ hyperons in neutrino-argon scatterings.

Furthermore, SBND’s large mass, its proximity to a high-intensity beam with high-statistics, and LArTPC's capabilities allow for the exploitation of a PRISM-like feature~\cite{2021APS..APRZ12005D}. 
This feature introduces an additional degree of freedom in performing targeted neutrino cross-section measurements by analyzing neutrino interactions across a continuous range of off-axis angles, corresponding to fluxes with mean energy shifts of $\sim$200 MeV. 
The SBND-PRISM ability of taking measurements in the same beam and with the same detector, but with different neutrino spectrum exposures by selecting events in different angular bins on the face of the detector, will be a powerful technique for disentangling nuclear effects in neutrino-argon interactions and constraining neutrino-nucleus models and generators.

SBND began collecting BNB physics data in December 2024 and is expected to operate with an exposure of $10\times10^{20}$ protons-on-target until the planned long accelerator shutdown at Fermilab in late 2027 or early 2028, yielding nearly 10 million neutrino charged-current and neutral-current interactions on argon. 
This data set will be instrumental in advancing our understanding of neutrino-argon scattering at the sub-GeV and GeV energy scales. After the accelerator expected restart in 2029, opportunities are being explored to operate SBND in antineutrino mode, addressing the scarcity of antineutrino-argon scattering data and its potential importance for DUNE's CP violation measurement. 
The insights of SBND data will play a critical role in refining neutrino-argon interaction models, providing essential inputs for DUNE and future neutrino experiments. 

MicroBooNE was the first operational LArTPC located 470 meters away from BNB. 
It consists of 85-ton active mass, ran for 5 years and collected $12.5\times10^{20}$ protons-on-target. 
Using both the BNB beamline and off-axis neutrinos from the Main Injector NuMI beam, MicroBooNE diverse cross section program supplied some of the highest-statistics neutrino interaction measurements, including multiple final states with different numbers of protons~\cite{MicroBooNE:2023tzj,MicroBooNE:2022emb,MicroBooNE:2020akw}, pions~\cite{MicroBooNE:2018neo} and rare process such as $\Lambda$ and $\eta$ production~\cite{MicroBooNE:2022cls, MicroBooNE:2023ubu}. 

In the meantime, ICARUS has transitioned from its early operations in Italy and subsequent refurbishment at CERN to FNAL. 
After completing commissioning in June 2022, it is now analyzing data from its initial two physics runs and preparing for a third. 
Although its primary mission as the far detector in the SBN program is to search for sterile neutrinos, ICARUS also supports 
detailed cross-section measurements.
Using both the BNB beamline and off-axis neutrinos from the Main Injector NuMI beam, the experiment captures neutrino interactions over a several-GeV range, a capability enhanced by its off-axis positioning and the contributions from pion and kaon decays producing excellent electron and muon neutrino beams. This energy window is critical, as it overlaps the regions relevant to both SBN oscillation studies and a portion of the DUNE spectrum.

\textbf{T2K/HK:} The next generation experiment Hyper-Kamiokande (HK) will study the oscillations of both atmospheric and beam neutrinos, measuring both the extent of CP violation in neutrino oscillations and the neutrino mass ordering (MO). 
As such, it is concerned with neutrino interactions in the range $\mathcal{O}(0.5-50~\text{GeV})$.
The neutrino energy spectra from these sources are identical to those observed and studied by the Tokai--to--Kamioka (T2K, beam neutrinos) and Super--Kamiokande (SK, atmospheric neutrinos) experiments. 
HK directly benefits from developments in current-generation analyses.
The sensitivity to CP violation comes primarily from measuring the $\numuany \rightarrow \nueany$ oscillations in the beam, but also from the atmospheric flavour-changing channels at energies below 1 GeV.
Studies of HK's sensitivities to neutrino oscillations demonstrate that improving the systematic uncertainty over T2K's 2020 analysis~\cite{PhysRevD.103.112008} can lead to less than halve the required running time to exclude CP conservation at $5\sigma$ for 50\% of the phase space. 
The largest impact comes from improving the $\nu_e/\bar{\nu}_e$ uncertainty, although improvements in low-energy modelling and single-pion production are also focal points~\cite{T2K:2023smv,T2K:2024wfn}. 
To better understand these processes, HK will use two complimentary near detectors: the upgraded ND280 detector~\cite{abe2020t2knd280upgrade}, capable of measuring protons with a 200-300 MeV/c threshold and neutrons, and resolving high-multiplicity events, and the Intermediate Water Cherenkov Detector (IWCD), which will make high-statistics measurements of interactions on water, with large electron neutrino and anti-neutrino samples. IWCD has the ability to probe neutrino energy by moving to different off-axis angles from the beam using the PRISM technique~\cite{nuPRISM:2014mzw}, and particle and kinematics acceptance closely matching the far detector.

For T2K and HK's baseline and neutrino energy spectrum, the CP violating phase and the normal ordering can be degenerate in certain parts of the phase space ($\delta_{CP}\sim\pi/2$ in normal ordering, $\delta_{CP}\sim-\pi/2$ in inverted ordering). 
These degeneracies are almost entirely lifted when atmospheric and beam neutrinos are analyzed simultaneously~\cite{T2K:2024wfn}, since the atmospheric neutrinos are sensitive to the effects in a markedly different way to beam neutrinos.
Like SK and other atmospheric experiments, HK's sensitivity to the mass ordering comes from measuring the resonance region from matter effects in the Earth at $E_\nu\sim\mathcal{O}(3-10~\text{GeV})$. 
Importantly, the effect depends on the sign of the interacting neutrino, which is impossible to accurately reconstruct on an event-by-event basis.
However, knowledge of how neutrinos and anti-neutrinos differ in their interaction can permit some, albeit imperfect, separation.
Thus, improvements in DIS modelling that dominates in the mass ordering-sensitive region, as well as how DIS interactions differ for neutrinos and anti-neutrinos, has a profound impact on the sensitivity to the mass ordering for atmospheric neutrinos.
Furthermore, as a result of the $\delta_{CP}-\text{MO}$ degeneracy, it also impacts HK's ability to measure $\delta_{CP}$ at high significance if Nature prefers the region of maximal degeneracy.

Although HK will not see $\nu_\tau$ charged-current (CC) interactions in the beam due to the neutrino energy threshold, it will in the atmospheric neutrino samples. 
SK has demonstrated statistical separation of such events to produce several $\tau$-enhanced samples~\cite{Super-Kamiokande:2012xtd,Super-Kamiokande:2017edb}. These samples are being used to both measure the absolute $\nu_\tau$ interaction, which is at $1-2\sigma$ contention between OPERA, IceCube and KM3NeT~\cite{Mandal:2024zxs}.
They further enhance the sensitivity to neutrino oscillations and testing of the three-flavour oscillation paradigm.
Notably, the $\nu_\tau$ cross section is a significant uncertainty for the mass ordering sensitivity, since the products of decaying $\tau$-leptons can be mistaken for hadronisation products from an electron or muon (anti-)neutrino partaking in a DIS interaction. 
These events can thereby enter the same samples as those sensitive to the mass ordering, constituting an irreducible background. 
An improved understanding of the $\nu_\tau$ interaction cross section would therefore also improve the sensitivity to the neutrino mass ordering.
Critically, whereas HK's near detectors will provide constraints on neutrino interactions in the CP-sensitive GeV-scale range, they will not be able to provide strong constraints on the higher energy DIS and $\nu_\tau$ interactions, which impact the mass ordering sensitivity.

\textbf{NOvA:} The NOvA long-baseline experiment studies neutrino oscillations using Fermilab’s NuMI beamline and two liquid scintillator detectors.
Its narrow-band energy spectrum, peaked at $\sim$2 GeV, optimizes sensitivity to electron neutrino appearance and muon neutrino disappearance at the Far Detector (FD). The near detector (ND) plays a central role in both oscillation and neutrino interaction studies.  
 A key component of NOvA’s oscillation analyses to date  is ``near-to-far extrapolation'', a method to predict FD event rates using ND observations while reducing dependence on theoretical models \cite{Acero:2019ksn,NOvA:2021nfi}.  
Beyond extrapolation, the ND high-statistics datasets
 can be used to measure neutrino cross sections in the 1–3 GeV energy range, where multiple interaction channels are relevant. NOvA sits in the transition region between quasielastic scattering, meson exchange currents, and resonance production, resulting in sensitivity to each of channels and the sum. The detector's fine segmentation enables precise tracking and calorimetry, allowing clear separation of muons, hadrons, and electromagnetic showers. With dedicated neutrino and anti-neutrino beam runs, the experiment can explore a wide range of inclusive and exclusive $\nu_e$ CC, $\nu_\mu$ CC, and NC interactions \cite{cremonesisnowmass2021}.

Neutrino interaction models consistently show discrepancies when compared to NOvA data \cite{NOvA:2020rbg,NOvA:2021eqi,NOvA:2022see, NOvA:2024zmr}, mirroring challenges seen in other experiments.  
Historically, the experiment has relied on GENIE as the default event generator~\cite{geniev3highlights}, tuned through targeted modifications informed by theoretical advancements, external data constraints, and ND observations, in order to refine the modeling of the different neutrino-nucleus reaction mechanisms.   
Systematic uncertainties are similarly managed through a combination of GENIE reweighting parameters and NOvA-specific adjustments. 
For future analyses, NOvA plans to adopt GENIE’s ``DUNE AR23'' Common Model Configuration (CMC) and \texttt{nusystematics}, benefiting from standardized cross-section variations and an expanded reweighting framework.  
These community-driven tools align with its current approach and will enhance consistency with other experiments.  
NOvA’s high-statistics data will help refine these models, and its carbon target provides a key comparison to future argon measurements, strengthening the framework for long-term use in oscillation and neutrino interaction studies.  

\textbf{DUNE/ProtoDUNE:} DUNE's full physics program consists of high-precision measurements of $\theta_{23}$, $\theta_{13}$, $\Delta m^2_{32}$, the neutrino mass ordering, and the CP violating phase $\delta_{\rm{CP}}$, as well as tests of the three-flavor paradigm as a way to search for new physics in neurino oscillations. 
Achieving these goals requires 600-1000 kt$\cdot$MW$\cdot$yr of data statistics, depending on the true values of the parameters. 
Not only this but similar to HK, it will also require systematics to be constrained at the few-percent level, which is beyond what is achieved in current experiments. 

The first priority of the experiment is to successfully deliver Phase I, consisting of 20 kt of far detector mass, a 1 MW beam and the Phase I near detector composed of a movable liquid argon detector (ND-LAr) with a downstream spectrometer (TMS) and an on-axis beam monitor (SAND). 
Full achievement of DUNE's physics program also requires Phase II. 
To reach the necessary statistical uncertainty, Phase II consists of 6-10 additional calendar years of operation with a greater than 2 MW beam and 40 kt of far detector mass. 
Additionally, the Phase II near detector is necessary to achieve sufficient systematic uncertainty control. The main component of the Phase II ND is a magnetized high-pressure gaseous argon time projection chamber surrounded by an electromagnetic calorimeter and a muon detector to the Phase I ND; upgrades to ND-LAr and SAND are also possible. The Phase II ND will constrain many aspects of DUNE's systematic uncertainty budget, but of particular relevance to this document are the unprecedented neutrino-argon interaction measurements that it will enable.

DUNE's wide band beam leads to particular requirements on the global neutrino-argon interaction program. The wide range of energies allows a full oscillation cycle to be observed but it also means that many different interaction topologies contribute to the observed events. Liquid and gas argon are the ideal detectors for these measurements. 
Both have exquisite capability to observe exclusive final states, with liquid argon allowing deployment at the many-kt scale. Gas argon by comparison cannot achieve high target mass but is magnetizable and has lower energy thresholds. The support of CERN has been key to the delivery of both technologies. Particularly, ProtoDUNE and the technical expertise provided by the Neutrino Platform have been instrumental in the testing of liquid argon technology. It has also delivered DUNE's first physics results, which are hadron-argon interaction measurements in the GeV energy regions directly applicable to DUNE. This role will continue to be crucial to demonstrate the readiness of Phase II technologies and to continue to provide further improved measurements of hadron-argon interactions. 
GeV-scale beam tests of gas argon time projection chambers (TPCs) have also been carried out at CERN and more are envisaged for the Phase II ND. Furthermore, deploying the Phase II far detectors requires cryostats that can contain them, with the Phase I cryostats having been provided by CERN. 

The knowledge of the relation between  true and reconstructed neutrino energy is critical to the DUNE program. 
In a liquid argon detector such as DUNE, a portion of the neutrino energy is reconstructed through visible energy deposits. 
However, a significant portion of the neutrino energy is transferred to particles which are not reconstructed, and bias the reconstructed neutrino energy. 
One of the main sources of missing energy are neutrons and low energy particles.
Since neutrons do not ionize the liquid argon, they cannot be detected by their tracks. 
Although some neutrons can be reconstructed via their secondary interactions, the cross section of these processes is not well known and relies on sub-nanosecond timing resolution in order to infer the kinematics of the neutron. 
The main mechanisms through which neutrons are produced in neutrino interactions is through final state interactions (FSI). 
In anti-neutrino interactions, in addition to those produced in FSI mechanisms, neutrons are also produced at the vertex. 

Another important source of missing energy is the production of charged pions.  
The latter deposit energy inside the detector through ionization, but it is not always possible to resolve the number of pions produced in an interaction. 
As a result, pion production processes contribute with integer multiples of the pion mass of 139.57 MeV to the neutrino energy bias. 
Charged pions are produced predominantly through resonant interactions, but also in shallow and deep inelastic scattering, as well as through FSI. 
In practice, neutrino experiments rely on the use of Monte Carlo generators to predict these quantities, and the latter rely on numerical hadronization routines and cascade models. 
The lack of theoretical guidance and experimental guidance in this region means that models targeting pion production process or hadronization have poor predictive power.  

Support for neutrino scattering theory and measurements in Europe is crucial for DUNE. DUNE has had significant contributions to its interaction systematic model development from European collaborators (e.g. CERN staff in EP-nu and theory) and is supportive of efforts to formalize and further this work, e.g. through a dedicated neutrino physics center at CERN.

\textbf{KM3NeT:} The KM3NeT neutrino infrastructure consists of two detectors, ORCA and ARCA~\cite{Adrian-Martinez:2016fdl}. 
ORCA is optimized for the study of atmospheric neutrinos with energies ranging from 3 GeV to several hundred GeV. Its primary objective is to determine the relative ordering of neutrino masses. This ordering affects neutrino oscillations in matter, leaving an imprint on the atmospheric neutrino flux through characteristic appearance/disappearance patterns for the different neutrino flavors as a function of energy and path through the Earth. Most of ORCA's sensitivity comes from upgoing events with energies between 5 and 10 GeV~\cite{KM3NeT:2021ozk,KM3NeT:2021rkn}. Therefore, precise knowledge of the atmospheric neutrino flux and neutrino-water cross section in this energy range is essential. Recent analyses with an initial six-line configuration have demonstrated ORCA’s immense potential not only for studying neutrino oscillations but also for probing BSM physics~\cite{KM3NeT:2024ecf,KM3NeT:2024jji,KM3NeT:2024pte,KM3NeT:2025tiq,KM3NeT:2025ftj}. The event generator used to simulate neutrino interaction in this analyses is GENIE.

ARCA’s main goal is the detection of high-energy neutrinos from astrophysical sources, with energies ranging from a few TeV to PeV~\cite{KM3NeT:2024paj}. Due to its geographical location and excellent pointing accuracy, ARCA is particularly well-suited for identifying neutrino sources within the Milky Way at TeV energies. Additionally, even with a partial configuration of 21 lines, ARCA recently reported the most energetic neutrino ever observed—at 220 PeV, opening a new window to ultra-high-energy neutrino astronomy~\cite{KM3NeT:2025npi}. At energies above few tens of TeV, neutrino interactions with matter become significant in a process known as Earth attenuation, where neutrinos can interact with quarks during their passage through the Earth before reaching the detector. These deep inelastic interactions depend on nuclear properties, making it essential to accurately model nuclear effects at high energies. Currently, the event generator used for ARCA simulations is the HEDIS module of GENIE~\cite{Garcia:2020jwr}, using the tune with the CSMS model~\cite{Cooper-Sarkar:2011jtt}.

\textbf{FASER$\nu$:} The FASER$\nu$ detector at CERN's LHC is poised to significantly advance our understanding of neutrino interactions and their potential to reveal new physics. Positioned 480 meters downstream from the ATLAS interaction point, FASER$\nu$ is uniquely situated to detect high-energy neutrinos produced in proton-proton collisions. A critical aspect of FASER$\nu$'s research program is the precise study of neutrino interactions across all three flavors—electron, muon, and $\tau$—at energies from few hundred GeV to multiple TeV. These measurements are essential for several reasons.
These include probing Parton Distribution Functions (PDFs) where high-energy neutrino interactions provide insights into the internal structure of protons and neutrons. 
Accurate cross-section measurements help refine PDFs, thereby providing crucial input to numerous measurements in high-energy physics and astroparticle physics.
Furthermore, the highly energetic neutrino interactions with the Tungsten target result in a DIS from a nucleon, and the resulting QCD-jet propagates through the nuclear medium, modifying the hadronic states observed e.g. by widening the angular scale of hadron distribution, and the emission of baryons from the nucleus. The precise resolution of the FASER$\nu$ experiment can provide important information on the the interactions between the nuclear matter and the original QCD jet.



\textbf{IceCube:} The IceCube Neutrino Observatory, located at the South Pole, instruments a cubic kilometer of ice with 5160 digital optical modules — each housing a 10" photomultiplier tube — arranged along 86 vertical strings. Its central DeepCore subarray lowers the energy threshold to capture atmospheric neutrinos above 5 GeV for studies of neutrino properties, interactions, and physics beyond the Standard Model, while the remaining 78 strings, optimized for a threshold near 100 GeV, target cosmic-ray and astrophysical neutrino sources.

Advances in detector modeling \cite{IceCubeCollaboration:2023wtb} and machine learning event reconstruction have enabled IceCube to achieve leading measurements of the oscillation parameters $\theta_{23}$ and $\Delta m^{2}_{32}$ via the $\nu_\mu$ disappearance channel using only atmospheric neutrinos \cite{PhysRevLett.134.091801}. It has also evaluated PMNS matrix unitarity by measuring $\nu_\tau$ appearance \cite{IceCube:2019dqi}, and its broad energy coverage supports diverse tests of non-standard neutrino interactions and phenomena beyond the Standard Model \cite{IceCubeCollaboration:2024nle}.
IceCube further probes neutrino interactions directly by analyzing the inelasticity distribution in the $50\,\text{GeV} < E_\nu < 560\,\text{GeV}$ range \cite{IceCube:2025tvs} and at TeV energies \cite{IceCube:2018pgc}, and by determining cross sections through the absorption of atmospheric neutrinos traversing Earth \cite{PhysRevD.104.022001}. The identification of flavor from interactions of astrophysical neutrinos benefits from future improvements in modeling accuracy of neutrino interactions at TeV and PeV, as exemplified by the detection of a Glashow Resonance event near 6.3 PeV \cite{IceCube:2021rpz} — a direct probe of astrophysical neutrino flavor composition — and by tests of quantum gravity \cite{ICECUBE:2023gdv}.  These results complement efforts by \textit{FASER} at the LHC and will benefit from the planned \textit{Forward Physics Facility}.

Scheduled to begin data taking in 2026, the IceCube Upgrade is the first step toward the larger IceCube-Gen2 project. This phase will add seven densely instrumented strings within DeepCore, lowering the detection threshold into the GeV range and reducing systematic uncertainties across all energies through dedicated calibration hardware \cite{Yanez}. The upgrade aims to produce leading measurements of atmospheric neutrino oscillation parameters, achieve world-leading precision in $\nu_\tau$ appearance measurements, and significantly advance our understanding of the neutrino mass ordering.  Because uncertainties in cross sections, detector response, and atmospheric fluxes all equally limit precision, reducing them simultaneously is essential. Hence, advancing our understanding of neutrino interaction physics is crucial to fully harnessing IceCube’s contributions to fundamental physics.




\textbf{NuSTORM:} Muon decay is one of the very few sources of electron neutrinos in the energy range, 0.1-10\,GeV, relevant for oscillation and long-baseline experiments. 
Muons have a sufficient lifetime to allow effective formation of pure beams. Therefore, the concept of storing muon beams for neutrino production has been explored in many different contexts~\cite{Neuffer:1983xya,Geer:1997iz,nuSTORM:2012jbd}. 
Using these beams for long-baseline experiments would require very high luminosity; this, in turn, makes the muon production system complex and expensive. 
On the other hand, for measuring cross sections (or in general fixed-target experiments), relatively simply muon production methods can be used based on existing proton beam facilities existent at many accelerator laboratories like FNAL and CERN. 
The storage rings for energies in the GeV range are compact as well. Beam instrumentation allows to know the neutrino flux at the percent-level and potentially much better than that. 

The opportunity to implement nuSTORM at CERN has been studied in detail~\cite{Ahdida:2020whw}.  The proposed facility is capable of storing $\mu^\pm$ beams with momentum of between 1\,GeV/c and 6\,GeV/c and a momentum spread of $\pm 16$\%. 
This will allow neutrino-scattering measurements to be made over the kinematic range of interest to the DUNE and Hyper-K collaborations.  Synthetic neutrino beams 
can be created by combining neutrinos from muon beams stored at a variety of different energies~\cite{Jurj:2025nmm}.

A detector placed on the axis of the nuSTORM production straight will receive a bright flash of muon neutrinos from pion decay followed by a series of pulses of muon and electron neutrinos from subsequent turns of the muon beam.  Appropriate instrumentation in the decay ring and production straight will be capable of determining the integrated neutrino flux with a precision of $\lesssim 1$\%.  The flavour composition of the neutrino beam from muon decay is known and the neutrino-energy spectrum can be calculated precisely using the Michel parameters and the optics of the muon decay ring.  The pion and muon momenta ($p_\pi$ and $p_\mu$) can be optimised to measure $\parenbar{\nu}_e-A$ and $\parenbar{\nu}_\mu-A$ interactions with percent-level precision over the neutrino-energy range $0.3 \lesssim E_\nu \lesssim 5.5$\,GeV and to search for light sterile neutrinos with excellent sensitivity.

Once the neutrino source is known at this level, the challenge for cross section measurements then becomes to have detector systems that can match that precision, covering a large phase space efficiently and reconstructing particle properties precisely.
This is a common challenge for all neutrino sources. 

A relatively less explored idea, but considered in nuSTORM feasibility studies at CERN, is to use these neutrino precision beam facilites as laboratories for new physics searches, especially for light dark sectors~\cite{nuSTORM:2022div}.



\textbf{Electron scattering:} An effective way to enhance our understanding of neutrino-nucleus interactions is through
electron-nucleus scattering experiments~\cite{e4nu,RevModPhys.80.189}.
Electrons share with neutrinos the vector component of the interaction with nuclei,
providing a clean and well-controlled probe of complex nuclear dynamics.
Electron beams can be precisely tuned to a well-defined energy, allowing for the
separation of different final states and enabling high-statistics experiments.
Electron-nucleus scattering can provide valuable insights into nuclear effects that
also influence neutrino interactions.
These include the nuclear structure and responses, since electron scattering experiments allow precise mapping of nucleon momentum distributions and nuclear densities and this information helps refine nuclear models used in neutrino interaction simulations.
Furthermore, FSI after a neutrino interacts with a nucleus lead to particles that propagate through the nuclear medium and experience rescattering, absorption, or emission.
With electron scattering, FSI can be studied selecting precise final states.
Additionally, different effects corresponding to different energy ranges
(quasielastic, resonance production, and deep-inelastic scattering) contribute to the total cross section in neutrino experiments.
Data provided by electron scattering serve as benchmarks for validating models across these regimes.
Furthermore, one of the key challenges in neutrino experiments is reconstructing the incoming neutrino energy from the observation of particles in the final state.
Electron scattering experiments provide a test for energy reconstruction methods by comparing 
the known beam energy with the one reconstructed in the way it is done in neutrino experiments.

Present electron scattering facilities pursuing an electron-nucleus scattering program aimed at neutrino
physics are the Thomas Jefferson National Accelerator Facility in the USA~\cite{mariani,JLab_Inclusive,JLab_Ti,JLab_Exclusive} (up to 12~GeV beam energy)
and the MAMI facility in Germany~\cite{MAMI1,MAMI2,Mihovilovic} (up to 1.6~GeV beam energy). In the near future, more experimental efforts from other experiments are planned (LDMX~\cite{LDMX} experiment, eALBA at the ALBA synchrotron in Barcelona, Spain)
or under construction (MESA, 155~MeV energy-recovery accelerator~\cite{MESAscience} in Mainz, Germany).


\section{Theory of $\nu$-nucleus scattering}
\begin{figure}[htb]
\begin{center}
\includegraphics[width=0.75\textwidth]{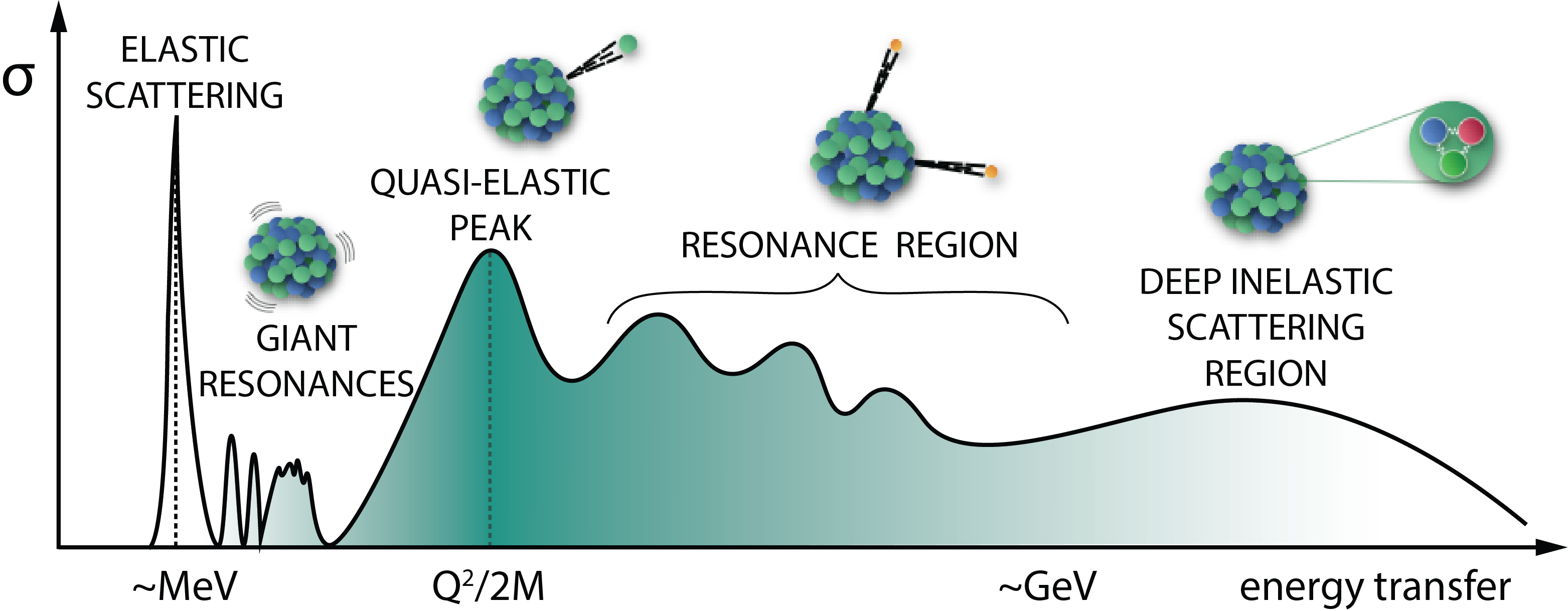}
\caption{Schematic view of $\nu$-nucleus scattering mechanisms.} 
\label{fig:mechanisms}
\end{center}
\end{figure}

Current and next-generation experiments studying and/or relying on  neutrino scattering processes for their physics programs span a vast energy range, from keV to TeV. Depending on the energy and momentum transfers, the scattered lepton (either a neutrino or a charged lepton) can excite various nuclear or hadronic final states. A schematic representation of the relevant interaction mechanisms is shown in Fig.~\ref{fig:mechanisms}.

Accurate cross-section calculations are essential in the era of precision neutrino physics, particularly when aiming at discovering new physics. However, these calculations are subject to theoretical uncertainties driven by nuclear and hadronic physics. A major challenge lies in the transition regions between different energy scales, where different degrees of freedom come into play. Close collaboration among nuclear, hadron, and particle physicists is required to treat these transitions reliably. Furthermore, in neutrino interactions,  both vector and axial currents come into play, but the latter remains poorly constrained. 

In this section, we provide a concise overview of the current status, challenges, and opportunities in the theory of neutrino cross sections across different energy scales.

\textbf{Coherent elastic neutrino-nucleus scattering:} \cevns is a low-energy neutrino interaction process in which a neutrino scatters off \textit{coherently} with an entire atomic nucleus, leaving it in its ground state~\cite{Freedman:1973yd}. One of the key advantages of \cevns is its large cross section compared to other neutrino interactions at low energies since, in the SM, it approximately scales quadratically with the number of neutrons. 
One of the key ingredients of the \cevns cross section is the nuclear form factor, which accounts for the finite size and internal structure of the nucleus, reflecting that the scattering process takes place on a many-body system. Its dependence on the momentum transfer is particularly relevant for experiments using neutrinos with relatively high energies, such as those using spallation sources. 
Detailed calculations are needed in order to model such a dependence, which may also be affected by new physics~\cite{Hoferichter:2020osn}. 
In the upcoming years, new data is expected from a plethora of \cevns experiments using more powerful neutrino sources, different nuclear targets, and improved detectors, aiming at more precise measurements and a significant reduction of systematic uncertainties. 
Thus, improving current theoretical models of nuclear form factors is crucial to boost the potential of \cevns experiments to search for BSM signals.

\textbf{Inelastic scattering up to and including the quasi-elastic peak:} At low and intermediate energy transfers of $\mathcal{O}(10)-\mathcal{O}(100)$ MeV, neutrino can excite low-lying nuclear resonances, up to the so-called quasi-elastic region in which the electroweak probe resolves individual nucleons. 
While the quasi-elastic peak at the few-hundreds MeV momentum transfer has been extensively studied in electron-scattering experiments (theoretical uncertainties and covariance matrices~\cite{PhysRevD.102.074012} for inclusive cross-sections are at the order of 10\%), the details of nuclear structure, upon which the description of specific exclusive channels depend, are far less constrained, especially in nuclei of relevance for future experiments such as argon and oxygen.
Moreover, the axial current contribution depends on nucleon form-factors which have to be either measured or calculated~\cite{PhysRevD.93.113015} from first principles. 
The interplay between one-nucleon and multi-nucleon knock-out processes, as well as the role of $\Delta(1232)$ excitation, are relevant topics of current research.
In this region the \textit{ab initio} nuclear calculations of several nuclei of interest have been recently performed~\cite{Ruso:2022qes}, delivering the state-of-the-art predictions together with the estimation of nuclear uncertainties. Other approximated approaches have also been proposed, employing the Hartree-Fock method, relativistic mean field, spectral functions or the short-time approximation.
The region of $\mathcal{O}(10)$ MeV is also challenging. The available calculations, based on the shell-model, or involving random-phase approximation,  still deliver predictions with unquantified uncertainties. 
This region is of relevance, in particular, for the correct interpretation of the signal from supernovae neutrinos, expected to be observed by the HK and DUNE detectors. Quasi-elastic scattering is the dominant mechanism for the T2K experiment (and future HK) and contributes to $\sim30\%$ of the event rates for DUNE. A failure to describe precisely the quasi-elastic signal can lead to a biased extraction of the neutrino oscillation parameters.

\textbf{Resonance region:} At higher energy transfers, electroweak interactions produce one or more mesons in the final state, e.g., $\pi N$, $\pi\pi N$, $\eta N$, $K N$. Associated strangeness production, leading to $K Y$ states with $Y = \Lambda, \Sigma, \dots$, also takes place.
%
Among these processes, single pion production stands out for its crucial role in the energy range relevant to accelerator and atmospheric neutrino experiments, acting as either a key signal component or as  a critical background that requires precise constraints. In spite of the significant recent progress, uncertainties 
persist in modeling pion production.

The description of meson production in the resonance region is done in terms of the effective hadronic degrees of freedom of QCD in the non-perturbative regime.  Close to threshold, Chiral Perturbation Theory allows for a systematic improvement by computing higher-order corrections, but a theoretical description covering the whole kinematic range probed by few-GeV neutrinos demands phenomenological modeling using external non-neutrino data. Indeed, thanks to approximate flavor symmetries and the partial conservation
of the axial current (PCAC), electron- and meson-nucleon scattering provide very valuable input for the description of weak inelastic processes. In particular, electron scattering data on hydrogen and deuterium allow the  isovector and isoscalar contributions to the vector current to be disentangled. On the other hand, PCAC relates the axial current at zero four-momentum transferred squared ($Q^2=0$) to the pion-nucleon elastic scattering amplitude (Goldberger-Treiman relations). However,  the axial current remains largely unconstrained at $Q^2 \neq 0$. For the most straightforward validation of the axial current, one currently has to rely on flux-averaged cross sections obtained in bubble chambers close to half a century ago~\cite{CC-ANL82_long, CC-BNL86, ALLASIA1990}.
The bubble chamber data are limited by statistics, and significant uncertainties due to deuteron final-state interactions~\cite{Nakamura:deuteronANL} and the knowledge of the absolute flux are not fully under control~\cite{NuSTEC:2020nsl}.
Modern measurements of neutrino-induced meson production on hydrogen or deuterium would be priceless to set stronger constraints~\cite{Snowmass2021:nu-H, Alvarez-Ruso:2022exy}. 

The description of pion production on nuclei is further complicated by many-body physics. Significant theoretical efforts towards the description of neutrino-induced pion production on nuclei have been reported.
Some effects that should be taken into account include nuclear momentum distributions, final-state interactions, and possible modification of resonance properties due to in-medium effects. The latter are intimately connected to meson-exchange currents in the nuclear medium, where the $\Delta$ resonance plays a crucial role.


\textbf{Shallow and deep inelastic scattering (SIS-DIS):} In the few GeV energy region, it is important to understand the transition region from the regime in which hadrons are the effective degrees of freedom, to the one of deep inelastic scattering (DIS), where the dynamics of quarks and gluons is described by perturbative QCD~\cite{SajjadAthar:2022pjt}. 
This transition region is commonly known as the shallow inelastic scattering (SIS) region. However, defining the exact kinematic boundaries between SIS and DIS is challenging because it is poorly understood both theoretically and experimentally. 

Extensive experimental studies in electron/muon-nucleon (e/$\mu$-N) scattering have been conducted in the SIS region, serving as a test bed for quark-hadron duality. 
In contrast, experimental measurements of electron/muon-nucleus scattering are scarce. Due to the lack of high-statistics data across the SIS region, studies of neutrino-nucleon and neutrino-nucleus scattering duality have relied primarily on theoretical models. Even these theoretical investigations are limited, with only a few comprehensive studies available in the literature\cite{Lalakulich:2006yn,Lalakulich:2009zza}. 
As more precise inclusive lepton scattering data from nucleons and nuclei become available, a rigorous verification of quark-hadron duality across the SIS region could offer a unified description of lepton-nucleon and lepton-nucleus interactions.

In the kinematical regime of DIS, the scattering cross section can be expressed in terms of nuclear structure functions, which are derived using QCD factorization theorems. In the weak sector, these structure functions have been primarily investigated by two groups~\cite{Kulagin:2007ju,Kulagin:2004ie,Zaidi:2019asc,Zaidi:2019mfd,Haider:2016zrk,Haider:2015vea,Ansari:2021cao}. 
Nucleon structure functions serve as the essential inputs for determining both the nuclear structure functions and the scattering cross section. Thus, a comprehensive understanding of nucleon structure functions is also crucial~\cite{Zaidi:2019asc,Ansari:2020xne}.
Nuclear effects of charged current deep inelastic $\nu/\nubar$-A scattering are also studied by the nuclear CTEQ (nCTEQ) collaboration by using data from 
the NuTeV, CCFR, CDHSW experiments on $\nu/\nubar$-Fe and CHORUS on $\nu/\nubar$-Pb.   The data that will be collected  at the upcoming neutrino experiments at the LHC will be fundamental in shedding light on the discrepancy between the correction factors  to the nuclear structure 
functions extracted in the neutrino and charged lepton scattering case.


\textbf{Lattice Quantum Chromodynamics:} LQCD offers an alternative strategy, complementary to existing experimental measurements, that leverages high performance supercomputing resources to provide insight about weak interactions with nuclear matter~\cite{USQCD:2022mmc,Ruso:2022qes}. 
%
Results from LQCD have already made an impact on our understanding of neutrino-nucleon scattering matrix elements.
Computations of quasielastic (QE) scattering with complete error budgets
 predicted that the QE cross section should be enhanced by as much as 30\%
 \cite{%
 RQCD:2019jai,
 Park:2021ypf,Djukanovic:2022wru,Meyer:2022mix,
 Jang:2023zts,Tomalak:2023pdi,
 Alexandrou:2023qbg,Tsuji:2023llh}.
Calculations of nucleon transition amplitudes in the first resonance region
 are already underway and limited first results should appear sometime within
 the next few years~\cite{%
 Alexandrou:2006mc,Barca:2022uhi,Barca:2024sub,
 Alexandrou:2024tin,Alexandrou:2024tps,Hackl:2024whw}.
These calculations rely on interpolating nucleon-pion states with explicit 3-momentum.
The theoretical and computational complexity of performing these calculations
 above the $N\pi\pi$ threshold is likely to be prohibitively challenging~\cite{%
 Blanton:2019igq,Briceno:2020vgp},
 so first results will likely be obtained from computations with unphysically heavy pion masses.
Alternative methods employing an optical theorem-like approach can provide
 estimates of hadronic matrix elements in the shallow inelastic scattering (SIS)
 region of phase space~\cite{%
 Liang:2019frk,Fukaya:2020wpp,Liang:2023uai}.
The use of LQCD for SIS is limited by working in Euclidean space to avoid a sign problem, since converting the Euclidean space response to one in Minkowski space is an ill-posed problem~\cite{%
 Hansen:2019idp,
 Horak:2021syv,
 Bulava:2021fre,
 DelDebbio:2024sfa}
 (but with bounded uncertainty~\cite{%
 Jay:2025dzl}).
LQCD computations can also be carried out to indirectly extract parton distribution functions, generalized distribution functions, and transverse momentum dependent distributions including those with axial matrix elements~\cite{%
 Ji:2013dva,
 Ma:2014jla,
 Radyushkin:2016hsy,
 Chambers:2017dov,
 Ji:2020ect,
 Constantinou:2020pek,
 Egerer:2021ymv,
 Bhattacharya:2023jsc,
 Delmar:2023agv,
 Gao:2023ktu
 }.

LQCD computations of hadron scattering also tell us about nuclear interactions responsible for binding into nuclei. Calculations of pion-nucleon~\cite{%
 Silvi:2021uya,Pittler:2021bqw,Bulava:2022vpq,Alexandrou:2023elk}
 and nucleon-nucleon~\cite{%
 Inoue:2011ai,Berkowitz:2015eaa,Iritani:2016jie,Wagman:2017tmp,
 Iritani:2018vfn,Drischler:2019xuo,Horz:2020zvv,Amarasinghe:2021lqa} spectra can be mapped onto hadronic scattering phase shifts. Fitting these phase shifts in an effective theory provides constraints on low energy constants, which can be used as inputs to nuclear ab initio and many body calculations. From these, information about intermediate nuclear states can be extracted for use in Monte Carlo generators.

\textbf{Interplay between cross-section determination and BSM searches:}
Accurate predictions of neutrino–nucleus interactions across different energy scales are essential for both uncovering and potentially obscuring signals of BSM physics. Inaccurate modeling of reaction mechanisms may hinder our ability to effectively probe BSM phenomena. In what follows, we discuss several BSM searches where a precise determination of neutrino cross sections plays a critical role.

{\it Low energy regime:} \cevns has emerged as a powerful tool for probing the existence of physics beyond the SM. The data collected using neutrinos from spallation sources and reactors has been vastly used to search for new physics, including non-standard neutrino interactions, sterile neutrinos, light dark matter, and new force carriers such as $Z'$ bosons among others~\cite{Arguelles:2022tki,Papoulias:2017qdn}. 
The observation of \cevns has been crucial to reject degeneracies present in oscillation data when new neutrino interactions are considered in the form of effective operators~\cite{Coloma:2017ncl, Coloma:2023ixt}. 
Additionally, the low-momentum transfer involved in \cevns makes it a key process to search for new neutrino interactions involving very light and feebly-interacting mediators, which are otherwise difficult to probe in neutrino scattering experiments~\cite{Farzan:2015doa}.
\cevns data has also been used to set new limits on neutrino magnetic moments, charge radii, neutrino millicharges, and to experimentally determine the weak mixing angle at very low momentum transfer~\cite{Abdullah:2022zue}. 
Very recently, the observation of \cevns for solar neutrinos has also been reported by dark matter direct detection experiments, and has already been used to search for new physics in the neutrino sector~\cite{AristizabalSierra:2024nwf, DeRomeri:2024iaw, Blanco-Mas:2024ale}. 
Additionally, \cevns is highly relevant in astrophysics, as it plays a crucial role in supernovae dynamics and neutron star cooling processes.

{\it Moderate energy regime: } The broad energy and baseline coverage of atmospheric neutrinos makes them ideal for exploring various physics scenarios.  In the three-neutrino ($3\nu$) mixing framework, sub-GeV neutrinos help probe the CP-violation phase~\cite{Akhmedov:2008qt}, making the study of neutrino quasi-elastic interactions essential~\cite{Arguelles:2022hrt}. Around the GeV scale, with neutrino scattering dominated by baryon resonance production, the matter effects in the neutrino propagation enable to resolve the mass ordering~\cite{Akhmedov:2006hb,Ribordy:2013xea,Olavarrieta:2024eaq}. At higher energies, interactions are dominated by deep inelastic scattering, while flavor oscillations are controlled by $\Delta m^2_{31}$ and $\sin^2\theta_{23}$. 

A challenging but promising approach would constrain BSM non-standard or generalized interactions by combining measurements of the neutrino cross sections on hydrogen and deuterium for specific processes with precise and reliable predictions from LQCD used as model independent and BSM-free input for the hadronic tensor. Such a strategy, currently applied in new-physics searches from hadronic decays and flavor anomalies, has been explored for quasielastic and elastic neutrino-nucleon interactions in recent studies~\cite{Tomalak:2024yvq,Ilma:2024lkp}. 

Beyond the $3\nu$ scenario, atmospheric neutrinos offer a way to explore other BSM physics. For instance, Heavy Neutral Leptons (HNLs), that could explain the smallness of neutrino masses, couple to active states via the lepton mixing matrix ($\nu_{\alpha L} = \sum^{3}_{i=0} U_{\alpha i}\nu_{iL} + U_{\alpha 4}N^{c}_{4 R}$). HNLs can be produced in neutral current interactions and be detected through their decay into charged particles due to their coupling to active neutrinos~\cite{Coloma:2017ppo,Atkinson:2021rnp,Gustafson:2022rsz}. The analysis of 10 years of data by IceCube~\cite{IceCube:2025kve}, has reached a sensitivity of $U^2_{\tau 4}\sim 0.1$.
HNLs might also couple to active neutrinos through higher-dimensional operators such as the transition dipole moment, described at low energies by the effective operator $\mathcal{L} \supset \mu_{\text{tr}}\overline{\nu}{\alpha L}\sigma{\rho \sigma} N_{4R} F^{\rho\sigma}$, where, $F^{\rho\sigma}$ is the electromagnetic tensor. The neutrino-nucleon scattering is then mediated via photon exchange. For HNL masses around the GeV scale~\cite{Coloma:2017ppo,Atkinson:2021rnp,Gustafson:2022rsz}, it is possible to reach $\mu_{\text{tr}} \sim 10^{-10}\mu_{B}$.

{\it High energy regime:}
FASER$\nu$'s exploration of high-energy neutrino interactions not only tests the SM predictions but also serves as a powerful tool to uncover potential new physics. The detector's unique position and capabilities make it a pivotal experiment in advancing our understanding of fundamental particles and their interactions.

Neutrinos detected by FASER$\nu$ are primarily produced from the decays of hadrons such as pions, kaons, and charmed mesons. Studying these neutrinos enhances our understanding of forward particle production in high-energy collisions in the laboratory, as well as of high-energy cosmic ray interactions in the atmosphere.
On top of that, by comparing interaction rates among different neutrino flavors, FASER$\nu$ can assess whether the fundamental interactions are consistent across lepton flavors, as predicted by the SM. Deviations from expected ratios could indicate new physics phenomena.
Finally, recent studies have highlighted the capabilities of FASER$\nu$ in probing physics beyond the SM in neutrino interactions. In the framework of the Standard Model Effective Field Theory (SMEFT) and its low-energy counterpart, the Weak Effective Field Theory (WEFT), it has been shown that FASER$\nu$ will constrain interactions that are two to three orders of magnitude weaker than the SM weak interactions, effectively probing new physics at the multi-TeV scale. In certain scenarios, the constraints provided by FASER$\nu$ could be comparable to, or even surpass, existing limits from rare meson decays and other low-energy probes.





\section{Event Generators}

European institutes and laboratories play a critical role in developing, supporting, and running all of the neutrino event generators in use today (GENIE~\cite{Andreopoulos:2009rq}, GiBUU~\cite{Buss:2011mx}, NuWro~\cite{Juszczak:2005zs,Golan:2012rfa}, NEUT~\cite{Hayato:2021heg}, ACHILLES~\cite{Isaacson:2022cwh}, FLUKA/NUNDIS~\cite{Battistoni:2009jen}). 
With the advent of precision long-baseline experiments (DUNE, HK), large-scale atmospheric neutrino experiments and neutrino telescopes (KM3NeT, IceCube, HK), short-baseline experiments (SBND, ICARUS), collider-based neutrino experiments (FASERnu, SND@LHC), and sophisticated multi-experiment analyses (SK+T2K, T2K+NOvA), the scope of physics which the generators have to support is significantly expanding, as has their user base.

The range of neutrino energies needing modeling spans from 0.1 GeV to 1 TeV, while the nuclear targets vary from hydrogen and simple hydrocarbons to argon and even heavier materials like iron. Most existing neutrino event generators focus on the GeV energy scale—critical for accelerator-based oscillation experiments—but experimental needs extend beyond this range. Each energy regime comes with its own unique modeling challenges that must be addressed. For example, MARLEY is a dedicated neutrino event generator for the MeV scale and has been adopted for all official DUNE studies on supernova and solar neutrinos. 
The energy transfers of interest for such studies lie substantially below the quasielastic region, necessitating a distinct nuclear physics treatment from what is typical in neutrino event generators designed for higher energies. 
However, there is presently no dedicated support for development of MARLEY or a similar tool, and many features needed to execute real-world analyses with uncertainty quantification are currently missing. Fully exploiting the physics potential of the future neutrino program, including neutrino astrophysics and exotic physics searches at the MeV scale, will require additional investment in neutrino event generator development at ``nontraditional'' energies.

Existing data and generator analyses have demonstrated that no generator or theory is able to explain all current neutrino scattering data.
This has necessitated experiment-specific tuning, which often spoils the ability to reliably extrapolate in neutrino energy and target material.
The challenge of reliably modeling a large range of neutrino energy and interactions on a wide array of targets is particularly important for future joint oscillation analyses between experiments, which will provide the strongest constraints on the charge-parity (CP) violation and neutrino mass ordering with DUNE and HK. 
It is also central to experiments conducting analyses with neutrinos from different sources (e.g. HK's beam neutrinos combined with atmospheric neutrinos), and experiments with different nuclear targets (e.g. components of the SAND near detector in DUNE, or the upgraded ND280 near detector in Hyper-Kamiokande).

For the GeV-scale experiments to continue their success in the realm of high statistics, significant efforts are dedicated to reducing the systematic uncertainties from neutrino-nucleus interactions, which is often a dominant uncertainty. 
Oscillation analyses in T2K and NOvA see an impact of cross-section uncertainties on $\mathcal{O}(3-10\%)$, with statistical uncertainties of $\mathcal{O}(5-10\%)$. 
For the forthcoming experiments, the statistical uncertainties will drastically decrease to $\mathcal{O}(1-2\%)$, and the systematic uncertainties need to scale accordingly. 
Furthermore, neutrino interaction modelling may also bias the oscillation parameters, via interaction effects masquerading as oscillation effects.

Profound collaboration with the neutrino interaction community is necessary to continue addressing these issues. Furthermore, implementation of sophisticated neutrino interaction models also requires improvements in numerical methods, for instance sampling high dimensional phase spaces, and utilising and developing new machine learning techniques.
Efforts have also focused on providing a common event-generator format, NuHepMC~\cite{Gardiner:2023ejq}, which allows straightforward adoption of different neutrino interaction generators into production pipelines. The format has seen adoption by multiple generators and is being studied for deployment in experiments, such as HK.
Integration between the data and generator comparison and tuning framework NUISANCE~\cite{Stowell:2016jfr} and the collider-driven HepData~\cite{Maguire:2017ypu} initiative is also underway, to facilitate assessing shortcomings in generator modeling.
Cross-collaboration with collider-based efforts with the HEP software foundation (HSF) have also begun in earnest~\cite{Campbell:2022qmc}.

\vspace{2cm}
\noindent {\bf Acknowledgments}\\

\noindent The authors gratefully acknowledge Jorge Morfin, Ulrich Mosel, and Bei Zhou for valuable discussions concerning the broader context of the topics addressed in this document.  
We also thank all members of the NuSTEC collaboration, whose ongoing efforts and scientific exchanges have significantly contributed to the development of this work.
This work was performed under the auspices of the U.S. Department of Energy by Lawrence Livermore National Laboratory under Contract DE-AC52-07NA2734, the Neutrino Theory Network Program Grant DE-AC02-07CHI11359, and U.S. Department of Energy DE-SC0020250.

\clearpage


\bibliographystyle{unsrt}

\bibliography{epps_refs,refs}  

\begin{thebibliography}{100}

\bibitem{dune_1}
I.~Gil-Botella, S.~Gollapinni, C.~Marshall, S.~Soldner-Rembold, and M.~Sore.
\newblock {\textbf{The DUNE Science Program}}.
\newblock {- submission to the ESPP 2026 update, March 2025}.

\bibitem{dune_2}
I.~Gil-Botella, S.~Gollapinni, C.~Marshall, S.~Soldner-Rembold, and M.~Sore.
\newblock {\textbf{The DUNE Phase II Detectors}}.
\newblock {- submission to the ESPP 2026 update, March 2025}.

\bibitem{dune_3}
G.~Barker.
\newblock {\textbf{European Contributions to Fermilab Accelerator Upgrades and Facilities for the DUNE Experiment}}.
\newblock {- submission to the ESPP 2026 update, March 2025}.

\bibitem{dune_4}
J.~Bian, M.~Kirby, A.~McNab, and L.~Whitehead.
\newblock {\textbf{DUNE Software and Computing Research and Development}}.
\newblock {- submission to the ESPP 2026 update, March 2025}.

\bibitem{ESPP_hk}
{F. Di Lodovico et al}.
\newblock {\textbf{The Hyper-Kamiokande experiment: input to the update of the European Strategy for Particle Physics}}.
\newblock {- submission to the ESPP 2026 update, March 2025}.

\bibitem{ESPP_ESSnuSB}
G.~Fanourakis M.~Dracos, T.~Ekelöf.
\newblock {\textbf{Input from ESSnuSB (European Spallation Source neutrino Super Beam) to the 2026 update of the European Strategy for Particle Physics}}.
\newblock {- submission to the ESPP 2026 update, March 2025}.

\bibitem{ESSP_SBN}
F.~Terranova N.~Charitonidis, M. Perrin-Terrin.
\newblock {\textbf{{SBN}@{CERN}: A short-baseline neutrino beam at {CERN} for high-precision cross-section measurements}}.
\newblock {- submission to the ESPP 2026 update, March 2025, arXiv:2503.21589}.

\bibitem{ESSP_SBN_FNAL}
R.~Acciarri et~al.
\newblock {\textbf{The Short-Baseline Near Detector at Fermilab}}.
\newblock March 2025.

\bibitem{ESPP_t2k}
{K. Mahn et al}.
\newblock {\textbf{T2K Experiment: future plans and capabilities}}.
\newblock {- submission to the ESPP 2026 update, March 2025}.

\bibitem{ESSP_coherent}
K.~Scholberg J.~Hakenmueller.
\newblock {\textbf{The {COHERENT} Experiment}}.
\newblock {- submission to the ESPP 2026 update, March 2025}.

\bibitem{ESSP_nustorm}
{\textbf{Neutrinos from Stored Muons (nuSTORM)}}.
\newblock {- submission to the ESPP 2026 update, March 2025}.

\bibitem{ESSP_nu_theory}
M.~Escudero, J.~Kopp, M.~Ovchynnikov, and Z.~Tabrizi.
\newblock {\textbf{Neutrino Theory in the Precision Era}}.
\newblock {- submission to the ESPP 2026 update, March 2025}.

\bibitem{ESSP_nu_CERN}
F.~Terranova F.~Resnati, G.~Schnell.
\newblock {\textbf{Neutrinos@CERN: A Community-Driven Contribution to the Update of the European Strategy for Particle Physics}}.
\newblock {- submission to the ESPP 2026 update, March 2025}.

\bibitem{Nikolakopoulos:2019qcr}
Alexis Nikolakopoulos, Natalie Jachowicz, Nils Van~Dessel, Kajetan Niewczas, Ra\'ul Gonz\'alez-Jim\'enez, Jos\'e~Manuel Ud\'\i{}as, and Vishvas Pandey.
\newblock {Electron versus Muon Neutrino Induced Cross Sections in Charged Current Quasielastic Processes}.
\newblock {\em Phys. Rev. Lett.}, 123(5):052501, 2019.

\bibitem{PhysRevLett.129.081801}
D.~Akimov et~al.
\newblock Measurement of the coherent elastic neutrino-nucleus scattering cross section on csi by coherent.
\newblock {\em Phys. Rev. Lett.}, 129:081801, Aug 2022.

\bibitem{PhysRevLett.126.012002}
D.~Akimov et~al.
\newblock First measurement of coherent elastic neutrino-nucleus scattering on argon.
\newblock {\em Phys. Rev. Lett.}, 126:012002, Jan 2021.

\bibitem{adamski2024detectioncoherentelasticneutrinonucleus}
S.~Adamski et~al.
\newblock First detection of coherent elastic neutrino-nucleus scattering on germanium, 2024.

\bibitem{PhysRevLett.126.091301}
E.~Aprile et~al.
\newblock Search for coherent elastic scattering of solar $^{8}\mathrm{B}$ neutrinos in the xenon1t dark matter experiment.
\newblock {\em Phys. Rev. Lett.}, 126:091301, Mar 2021.

\bibitem{PhysRevLett.133.191002}
E.~Aprile et~al.
\newblock First indication of solar $^{8}\mathrm{B}$ neutrinos via coherent elastic neutrino-nucleus scattering with xenonnt.
\newblock {\em Phys. Rev. Lett.}, 133:191002, Nov 2024.

\bibitem{Ackermann:2025obx}
N.~Ackermann et~al.
\newblock {First observation of reactor antineutrinos by coherent scattering}.
\newblock {\em arXiv}, 1 2025.

\bibitem{Alekou:2022emd}
A.~Alekou et~al.
\newblock {The European Spallation Source neutrino super-beam conceptual design report}.
\newblock {\em Eur. Phys. J. ST}, 231(21):3779--3955, 2022.

\bibitem{Garoby:2017vew}
Roland Garoby et~al.
\newblock {The European Spallation Source Design}.
\newblock {\em Phys. Scripta}, 93(1):014001, 2018.

\bibitem{ESSnuSB:2023ogw}
A.~Alekou et~al.
\newblock {The ESSnuSB Design Study: Overview and Future Prospects}.
\newblock {\em Universe}, 9(8):347, 2023.

\bibitem{MicroBooNE:2015bmn}
R.~Acciarri et~al.
\newblock {A Proposal for a Three Detector Short-Baseline Neutrino Oscillation Program in the Fermilab Booster Neutrino Beam}.
\newblock {\em arXiv}, 3 2015.

\bibitem{2021APS..APRZ12005D}
Marco {Del Tutto}, Ornella {Palamara}, Vishvas {Pandey}, and {SBND Collaboration}.
\newblock {SBND-PRISM: Sampling Multiple Off-Axis Neutrino Fluxes with the Same Detector}.
\newblock In {\em APS April Meeting Abstracts}, volume 2021 of {\em APS Meeting Abstracts}, page Z12.005, January 2021.

\bibitem{MicroBooNE:2023tzj}
P.~Abratenko et~al.
\newblock {First Double-Differential Measurement of Kinematic Imbalance in Neutrino Interactions with the MicroBooNE Detector}.
\newblock {\em Phys. Rev. Lett.}, 131(10):101802, 2023.

\bibitem{MicroBooNE:2022emb}
P.~Abratenko et~al.
\newblock {First Measurement of Differential Cross Sections for Muon Neutrino Charged Current Interactions on Argon with a Two-proton Final State in the MicroBooNE Detector}.
\newblock 11 2022.

\bibitem{MicroBooNE:2020akw}
P.~Abratenko et~al.
\newblock {Measurement of differential cross sections for $\nu_{\mu}$ -Ar charged-current interactions with protons and no pions in the final state with the MicroBooNE detector}.
\newblock {\em Phys. Rev. D}, 102(11):112013, 2020.

\bibitem{MicroBooNE:2018neo}
C.~Adams et~al.
\newblock {First measurement of $\nu_{\mu}$ charged-current $\pi^{0}$ production on argon with the MicroBooNE detector}.
\newblock {\em Phys. Rev. D}, 99(9):091102, 2019.

\bibitem{MicroBooNE:2022cls}
P.~Abratenko et~al.
\newblock {First Measurement of Quasielastic \ensuremath{\Lambda} Baryon Production in Muon Antineutrino Interactions in the MicroBooNE Detector}.
\newblock {\em Phys. Rev. Lett.}, 130(23):231802, 2023.

\bibitem{MicroBooNE:2023ubu}
P.~Abratenko et~al.
\newblock {First Measurement of \ensuremath{\eta} Meson Production in Neutrino Interactions on Argon with MicroBooNE}.
\newblock {\em Phys. Rev. Lett.}, 132(15):151801, 2024.

\bibitem{PhysRevD.103.112008}
K.~Abe et~al.
\newblock Improved constraints on neutrino mixing from the t2k experiment with $3.13\ifmmode\times\else\texttimes\fi{}{10}^{21}$ protons on target.
\newblock {\em Phys. Rev. D}, 103:112008, Jun 2021.

\bibitem{T2K:2023smv}
K.~Abe et~al.
\newblock {Measurements of neutrino oscillation parameters from the T2K experiment using $3.6\times 10^{21}$ protons on target}.
\newblock {\em Eur. Phys. J. C}, 83(9):782, 2023.

\bibitem{T2K:2024wfn}
K.~Abe et~al.
\newblock {First Joint Oscillation Analysis of Super-Kamiokande Atmospheric and T2K Accelerator Neutrino Data}.
\newblock {\em Phys. Rev. Lett.}, 134(1):011801, 2025.

\bibitem{abe2020t2knd280upgrade}
K.~Abe et~al.
\newblock T2k nd280 upgrade -- technical design report, 2020.

\bibitem{nuPRISM:2014mzw}
S.~Bhadra et~al.
\newblock {Letter of Intent to Construct a nuPRISM Detector in the J-PARC Neutrino Beamline}, 12 2014.
\newblock {a}r{X}iv:1412.3086.

\bibitem{Super-Kamiokande:2012xtd}
K.~Abe et~al.
\newblock {Evidence for the Appearance of Atmospheric Tau Neutrinos in Super-Kamiokande}.
\newblock {\em Phys. Rev. Lett.}, 110(18):181802, 2013.

\bibitem{Super-Kamiokande:2017edb}
Z.~Li et~al.
\newblock {Measurement of the tau neutrino cross section in atmospheric neutrino oscillations with Super-Kamiokande}.
\newblock {\em Phys. Rev. D}, 98(5):052006, 2018.

\bibitem{Mandal:2024zxs}
Maitrayee Mandal.
\newblock {The appearance of tau neutrinos in the flux of atmospheric neutrinos at Super-Kamiokande}.
\newblock {\em PoS}, ICHEP2024:154, 2025.

\bibitem{Acero:2019ksn}
M.A. Acero et~al.
\newblock {First Measurement of Neutrino Oscillation Parameters using Neutrinos and Antineutrinos by NOvA}.
\newblock {\em Phys. Rev. Lett.}, 123(15):151803, 2019.

\bibitem{NOvA:2021nfi}
M.~A. Acero et~al.
\newblock {Improved measurement of neutrino oscillation parameters by the NOvA experiment}.
\newblock {\em Phys. Rev. D}, 106(3):032004, 2022.

\bibitem{cremonesisnowmass2021}
Linda Cremonesi, Mathew Muether, and Jonathan Paley.
\newblock {Snowmass 2021-Letter of Interest: The NOvA Near Detector Physics Program}, 2020.

\bibitem{NOvA:2020rbg}
M.~A. Acero et~al.
\newblock {Adjusting neutrino interaction models and evaluating uncertainties using NOvA near detector data}.
\newblock {\em Eur. Phys. J. C}, 80(12):1119, 2020.

\bibitem{NOvA:2021eqi}
M.~A. Acero et~al.
\newblock {Measurement of the double-differential muon-neutrino charged-current inclusive cross section in the NOvA near detector}.
\newblock {\em Phys. Rev. D}, 107(5):052011, 2023.

\bibitem{NOvA:2022see}
M.~A. Acero et~al.
\newblock {Measurement of the $\nu_e-$Nucleus Charged-Current Double-Differential Cross Section at $\left< E_{\nu} \right> = $ 2.4 GeV using NOvA}.
\newblock {\em Phys. Rev. Lett.}, 130(5):051802, 2023.

\bibitem{NOvA:2024zmr}
M.~A. Acero et~al.
\newblock {Measurement of d2\ensuremath{\sigma}/d|q\textrightarrow{}|dEavail in charged current \ensuremath{\nu}\ensuremath{\mu}-nucleus interactions at \ensuremath{\langle}E\ensuremath{\nu}\ensuremath{\rangle}=1.86\,\,GeV using the NOvA Near Detector}.
\newblock {\em Phys. Rev. D}, 111(5):052009, 2025.

\bibitem{geniev3highlights}
Luis Alvarez-Ruso et~al.
\newblock {Recent highlights from {GENIE} v3}.
\newblock {\em Eur. Phys. J. ST}, 230(24):4449--4467, 2021.

\bibitem{Adrian-Martinez:2016fdl}
S.~Adrian-Martinez et~al.
\newblock {Letter of intent for KM3NeT 2.0}.
\newblock {\em J. Phys. G}, 43(8):084001, 2016.

\bibitem{KM3NeT:2021ozk}
S.~Aiello et~al.
\newblock {Determining the neutrino mass ordering and oscillation parameters with KM3NeT/ORCA}.
\newblock {\em Eur. Phys. J. C}, 82(1):26, 2022.

\bibitem{KM3NeT:2021rkn}
S.~Aiello et~al.
\newblock {Combined sensitivity of JUNO and KM3NeT/ORCA to the neutrino mass ordering}.
\newblock {\em JHEP}, 03:055, 2022.

\bibitem{KM3NeT:2024ecf}
S.~Aiello et~al.
\newblock {Measurement of neutrino oscillation parameters with the first six detection units of KM3NeT/ORCA}.
\newblock {\em JHEP}, 10:206, 2024.

\bibitem{KM3NeT:2024jji}
S.~Aiello et~al.
\newblock {Search for quantum decoherence in neutrino oscillations with six detection units of KM3NeT/ORCA}.
\newblock {\em arXiv}, 10 2024.

\bibitem{KM3NeT:2024pte}
S.~Aiello et~al.
\newblock {Search for non-standard neutrino interactions with the first six detection units of KM3NeT/ORCA}.
\newblock {\em JCAP}, 02:073, 2025.

\bibitem{KM3NeT:2025tiq}
S.~Aiello et~al.
\newblock {Probing invisible neutrino decay with the first six detection units of KM3NeT/ORCA}.
\newblock {\em arXiv}, 1 2025.

\bibitem{KM3NeT:2025ftj}
S.~Aiello et~al.
\newblock {Study of tau neutrinos and non-unitary neutrino mixing with the first six detection units of KM3NeT/ORCA}.
\newblock {\em arXiv}, 2 2025.

\bibitem{KM3NeT:2024paj}
S.~Aiello et~al.
\newblock {Astronomy potential of KM3NeT/ARCA}.
\newblock {\em Eur. Phys. J. C}, 84(9):885, 2024.

\bibitem{KM3NeT:2025npi}
S.~Aiello et~al.
\newblock {Observation of an ultra-high-energy cosmic neutrino with KM3NeT}.
\newblock {\em Nature}, 638(8050):376--382, 2025.

\bibitem{Garcia:2020jwr}
Alfonso Garcia, Rhorry Gauld, Aart Heijboer, and Juan Rojo.
\newblock {Complete predictions for high-energy neutrino propagation in matter}.
\newblock {\em JCAP}, 09:025, 2020.

\bibitem{Cooper-Sarkar:2011jtt}
Amanda Cooper-Sarkar, Philipp Mertsch, and Subir Sarkar.
\newblock {The high energy neutrino cross-section in the Standard Model and its uncertainty}.
\newblock {\em JHEP}, 08:042, 2011.

\bibitem{IceCubeCollaboration:2023wtb}
R.~Abbasi et~al.
\newblock {Measurement of atmospheric neutrino mixing with improved IceCube DeepCore calibration and data processing}.
\newblock {\em Phys. Rev. D}, 108(1):012014, 2023.

\bibitem{PhysRevLett.134.091801}
R.~Abbasi et~al.
\newblock Measurement of atmospheric neutrino oscillation parameters using convolutional neural networks with 9.3 years of data in icecube deepcore.
\newblock {\em Phys. Rev. Lett.}, 134:091801, Mar 2025.

\bibitem{IceCube:2019dqi}
M.~G. Aartsen et~al.
\newblock {Measurement of Atmospheric Tau Neutrino Appearance with IceCube DeepCore}.
\newblock {\em Phys. Rev. D}, 99(3):032007, 2019.

\bibitem{IceCubeCollaboration:2024nle}
R.~Abbasi et~al.
\newblock {Search for an eV-Scale Sterile Neutrino Using Improved High-Energy \ensuremath{\nu}\ensuremath{\mu} Event Reconstruction in IceCube}.
\newblock {\em Phys. Rev. Lett.}, 133(20):201804, 2024.

\bibitem{IceCube:2025tvs}
R.~Abbasi et~al.
\newblock {Measurement of the inelasticity distribution of neutrino-nucleon interactions for $\mathbf{80~GeV<E_{\nu}<560~GeV}$ with IceCube DeepCore}.
\newblock {\em arXiv}, 2 2025.

\bibitem{IceCube:2018pgc}
M.~G. Aartsen et~al.
\newblock {Measurements using the inelasticity distribution of multi-TeV neutrino interactions in IceCube}.
\newblock {\em Phys. Rev. D}, 99(3):032004, 2019.

\bibitem{PhysRevD.104.022001}
R.~Abbasi et~al.
\newblock Measurement of the high-energy all-flavor neutrino-nucleon cross section with icecube.
\newblock {\em Phys. Rev. D}, 104:022001, Jul 2021.

\bibitem{IceCube:2021rpz}
M.~G. Aartsen et~al.
\newblock {Detection of a particle shower at the Glashow resonance with IceCube}.
\newblock {\em Nature}, 591(7849):220--224, 2021.
\newblock [Erratum: Nature 592, E11 (2021)].

\bibitem{ICECUBE:2023gdv}
R.~Abbasi et~al.
\newblock {Search for decoherence from quantum gravity with atmospheric neutrinos}.
\newblock {\em Nature Phys.}, 20(6):913--920, 2024.

\bibitem{Yanez}
Juan~Pablo Yanez.
\newblock A decade of atmospheric neutrino oscillations with icecube.
\newblock In {\em NEUTRINO 2024}. Zenodo, 2024.

\bibitem{Neuffer:1983xya}
David Neuffer.
\newblock {Principles and Applications of Muon Cooling}.
\newblock {\em Conf. Proc. C}, 830811:481--484, 1983.

\bibitem{Geer:1997iz}
S.~Geer.
\newblock {Neutrino beams from muon storage rings: Characteristics and physics potential}.
\newblock {\em Phys. Rev. D}, 57:6989--6997, 1998.
\newblock [Erratum: Phys.Rev.D 59, 039903 (1999)].

\bibitem{nuSTORM:2012jbd}
P.~Kyberd et~al.
\newblock {$\nu$STORM - Neutrinos from STORed Muons: Letter of Intent}.
\newblock {\em arXiv}, 6 2012.

\bibitem{Ahdida:2020whw}
C.~C. Ahdida et~al.
\newblock {nuSTORM at CERN: Feasibility Study}.
\newblock {\em CERN-PBC-REPORT-2019-003}, 10 2020.

\bibitem{Jurj:2025nmm}
Paul Jurj.
\newblock {nuSTORM: neutrino physics on the path to the muon collider}.
\newblock {\em PoS}, ICHEP2024:825, 2025.

\bibitem{nuSTORM:2022div}
L.~Alvarez Ruso et~al.
\newblock {Neutrinos from Stored Muons (nuSTORM)}.
\newblock In {\em {Snowmass 2021}}, 3 2022.

\bibitem{e4nu}
A.M. Ankowski et~al.
\newblock Electron scattering and neutrino physics.
\newblock {\em Journal of Physics G: Nuclear and Particle Physics}, 50(12):120501, 2023.

\bibitem{RevModPhys.80.189}
Omar Benhar, Donal Day, and Ingo Sick.
\newblock Inclusive quasielastic electron-nucleus scattering.
\newblock {\em Rev. Mod. Phys.}, 80:189--224, Jan 2008.

\bibitem{mariani}
H.~Dai et~al.
\newblock {First measurement of the $\mathrm{Ar}(e,e')X$ cross section at Jefferson Laboratory}.
\newblock {\em Phys. Rev. C}, 99:054608, May 2019.

\bibitem{JLab_Inclusive}
M.~Murphy et~al.
\newblock Measurement of the cross sections for inclusive electron scattering in the e12-14-012 experiment at jefferson lab.
\newblock {\em Phys. Rev. C}, 100:054606, Nov 2019.

\bibitem{JLab_Ti}
H.~Dai et~al.
\newblock First measurement of the $\mathrm{Ti}(e,{e}^{\ensuremath{'}})x$ cross section at jefferson lab.
\newblock {\em Phys. Rev. C}, 98:014617, Jul 2018.

\bibitem{JLab_Exclusive}
L.~Gu et~al.
\newblock Measurement of the $\mathrm{Ar}(e,{e}^{\ensuremath{'}}p)$ and $\mathrm{Ti}(e,{e}^{\ensuremath{'}}p)$ cross sections in jefferson lab hall a.
\newblock {\em Phys. Rev. C}, 103:034604, Mar 2021.

\bibitem{MAMI1}
H.~Herminghaus et~al.
\newblock {The design of a cascaded 800 MeV normal conducting C.W. race track microtron}.
\newblock {\em Nuclear Instruments and Methods}, 138(1):1--12, 1976.

\bibitem{MAMI2}
K.-H. Kaiser et~al.
\newblock {The 1.5 GeV harmonic double-sided microtron at Mainz University}.
\newblock {\em Nuclear Instruments and Methods in Physics Research Section A: Accelerators, Spectrometers, Detectors and Associated Equipment}, 593(3):159--170, 2008.

\bibitem{Mihovilovic}
M.~Mihovilovi\v{c} et~al.
\newblock {Measurement of the $\mathrm {{}^{12}C}(e,e')$ Cross Sections at $Q^2=0.8\,\textrm{GeV}^2/c^2$}.
\newblock {\em Few Body Syst.}, 65(3):78, 2024.

\bibitem{LDMX}
Torsten Åkesson, Asher Berlin, Nikita Blinov, Owen Colegrove, Giulia Collura, Valentina Dutta, Bertrand Echenard, Joshua Hiltbrand, David~G. Hitlin, Joseph Incandela, John Jaros, Robert Johnson, Gordan Krnjaic, Jeremiah Mans, Takashi Maruyama, Jeremy McCormick, Omar Moreno, Timothy Nelson, Gavin Niendorf, Reese Petersen, Ruth Pöttgen, Philip Schuster, Natalia Toro, Nhan Tran, and Andrew Whitbeck.
\newblock Light dark matter experiment (ldmx), 2018.

\bibitem{MESAscience}
Sören Schlimme et~al.
\newblock The mesa physics program, proceedings of the 16th edition of the international conference on meson-nucleon physics and the structure of the nucleon" (menu 2023), 2024.

\bibitem{Freedman:1973yd}
Daniel~Z. Freedman.
\newblock {Coherent Neutrino Nucleus Scattering as a Probe of the Weak Neutral Current}.
\newblock {\em Phys. Rev. D}, 9:1389--1392, 1974.

\bibitem{Hoferichter:2020osn}
Martin Hoferichter, Javier Men\'endez, and Achim Schwenk.
\newblock {Coherent elastic neutrino-nucleus scattering: EFT analysis and nuclear responses}.
\newblock {\em Phys. Rev. D}, 102(7):074018, 2020.

\bibitem{PhysRevD.102.074012}
Kaushik Borah, Richard~J. Hill, Gabriel Lee, and Oleksandr Tomalak.
\newblock Parametrization and applications of the low-${Q}^{2}$ nucleon vector form factors.
\newblock {\em Phys. Rev. D}, 102:074012, Oct 2020.

\bibitem{PhysRevD.93.113015}
Aaron~S. Meyer, Minerba Betancourt, Richard Gran, and Richard~J. Hill.
\newblock Deuterium target data for precision neutrino-nucleus cross sections.
\newblock {\em Phys. Rev. D}, 93:113015, Jun 2016.

\bibitem{Ruso:2022qes}
L.~Alvarez Ruso et~al.
\newblock Theoretical tools for neutrino scattering: interplay between lattice qcd, efts, nuclear physics, phenomenology, and neutrino event generators.
\newblock {\em arXiv}, Mar 2022.

\bibitem{CC-ANL82_long}
G.~M. Radecky, V.~E. Barnes, D.~D. Carmony, A.~F. Garfinkel, M.~Derrick, E.~Fernandez, L.~Hyman, G.~Levman, D.~Koetke, B.~Musgrave, P.~Schreiner, R.~Singer, A.~Snyder, S.~Toaff, S.~J. Barish, A.~Engler, R.~W. Kraemer, K.~Miller, B.~J. Stacey, R.~Ammar, D.~Coppage, D.~Day, R.~Davis, N.~Kwak, and R.~Stump.
\newblock Study of single-pion production by weak charged currents in low-energy $\ensuremath{\nu}d$ interactions.
\newblock {\em Phys. Rev. D}, 25:1161--1173, Mar 1982.

\bibitem{CC-BNL86}
T.~Kitagaki et~al.
\newblock Charged-current exclusive pion production in neutrino-deuterium interactions.
\newblock {\em Phys. Rev. D}, 34:2554--2565, Nov 1986.

\bibitem{ALLASIA1990}
D.~Allasia, C.~Angelini, G.W. {van Apeldoorn}, A.~Baldini, S.M. Barlag, L.~Bertanza, F.~Bobisut, P.~Capiluppi, P.H.A. {van Dam}, M.L. Faccini-Turluer, A.G. Frodesen, G.~Giacomelli, H.~Huzita, B.~Jongejans, G.~Mandrioli, A.~Marzari-Chiesa, R.~Pazzi, L.~Ramello, A.~Romero, A.M. Rossi, A.~Sconza, P.~Serra-Lugaresi, A.G. Tenner, and D.~Vignaud.
\newblock Investigation of exclusive channels in $\nu$/$\overline{\nu}$-deuteron charged current interactions.
\newblock {\em Nuclear Physics B}, 343(2):285--309, 1990.

\bibitem{Nakamura:deuteronANL}
S.~X. Nakamura, H.~Kamano, and T.~Sato.
\newblock Impact of final state interactions on neutrino-nucleon pion production cross sections extracted from neutrino-deuteron reaction data.
\newblock {\em Phys. Rev. D}, 99:031301, Feb 2019.

\bibitem{NuSTEC:2020nsl}
L.~Aliaga et~al.
\newblock {Summary of the NuSTEC Workshop on Neutrino-Nucleus Pion Production in the Resonance Region}.
\newblock In {\em {NuSTEC Workshop on Neutrino-Nucleus Pion Production in the Resonance Region}}, 11 2020.

\bibitem{Snowmass2021:nu-H}
R.~Hill, T.~Junk, et~al.
\newblock {Snowmass 2021 LoI: Neutrino Scattering Measurements on Hydrogen and Deuterium }.
\newblock {\em SNOWMASS2021}, 2020.

\bibitem{Alvarez-Ruso:2022exy}
L.~Alvarez-Ruso et~al.
\newblock {Bubble Chamber Detectors with Light Nuclear Targets: A Snowmass 2021 White Paper}.
\newblock {\em arXiv}, 3 2022.

\bibitem{SajjadAthar:2022pjt}
M.~Sajjad~Athar, A.~Fatima, and S.~K. Singh.
\newblock {Neutrinos and their interactions with matter}.
\newblock {\em Prog. Part. Nucl. Phys.}, 129:104019, 2023.

\bibitem{Lalakulich:2006yn}
O.~Lalakulich, W.~Melnitchouk, and E.~A. Paschos.
\newblock {Quark-hadron duality in neutrino scattering}.
\newblock {\em Phys. Rev. C}, 75:015202, 2007.

\bibitem{Lalakulich:2009zza}
O.~Lalakulich, C.~Praet, N.~Jachowicz, J.~Ryckebusch, T.~Leitner, O.~Buss, and U.~Mosel.
\newblock {Neutrinos and duality}.
\newblock {\em AIP Conf. Proc.}, 1189(1):276--282, 2009.

\bibitem{Kulagin:2007ju}
Sergey~A. Kulagin and R.~Petti.
\newblock {Neutrino inelastic scattering off nuclei}.
\newblock {\em Phys. Rev. D}, 76:094023, 2007.

\bibitem{Kulagin:2004ie}
Sergey~A. Kulagin and R.~Petti.
\newblock {Global study of nuclear structure functions}.
\newblock {\em Nucl. Phys. A}, 765:126--187, 2006.

\bibitem{Zaidi:2019asc}
F.~Zaidi, H.~Haider, M.~Sajjad~Athar, S.~K. Singh, and I.~Ruiz~Simo.
\newblock {Weak structure functions in $\nu_l-N$ and $\nu_l-A$ scattering with nonperturbative and higher order perturbative QCD effects}.
\newblock {\em Phys. Rev. D}, 101(3):033001, 2020.

\bibitem{Zaidi:2019mfd}
F.~Zaidi, H.~Haider, M.~Sajjad~Athar, S.~K. Singh, and I.~Ruiz~Simo.
\newblock {Nucleon and Nuclear Structure Functions with Nonperturbative and Higher Order Perturbative QCD Effects}.
\newblock {\em Phys. Rev. D}, 99(9):093011, 2019.

\bibitem{Haider:2016zrk}
H.~Haider, F.~Zaidi, M.~Sajjad~Athar, S.~K. Singh, and I.~Ruiz~Simo.
\newblock {Nuclear medium effects in $F_{2A}^{EM}(x,Q^2)$ and $F_{2A}^{Weak}(x,Q^2)$ structure functions}.
\newblock {\em Nucl. Phys. A}, 955:58--78, 2016.

\bibitem{Haider:2015vea}
H.~Haider, F.~Zaidi, M.~Sajjad~Athar, S.~K. Singh, and I.~Ruiz~Simo.
\newblock {Nuclear medium effects in structure functions of nucleon at moderate $Q^2$}.
\newblock {\em Nucl. Phys. A}, 943:58--82, 2015.

\bibitem{Ansari:2021cao}
V.~Ansari, M.~Sajjad Athar, H.~Haider, I.~Ruiz Simo, S.~K. Singh, and F.~Zaidi.
\newblock {Deep inelastic (anti)neutrino-nucleus scattering}.
\newblock {\em Eur. Phys. J Special Topics}, (hep-ph/2106.14670), 6 2021.

\bibitem{Ansari:2020xne}
V.~Ansari, M.~Sajjad~Athar, H.~Haider, S.~K. Singh, and F.~Zaidi.
\newblock {Nonperturbative and higher order perturbative effects in deep inelastic $\nu_\tau/\bar\nu_\tau-$nucleon scattering}.
\newblock {\em Phys. Rev. D}, 102(11):113007, 2020.

\bibitem{USQCD:2022mmc}
Andreas~S. Kronfeld et~al.
\newblock {Lattice QCD and Particle Physics}.
\newblock 7 2022.

\bibitem{RQCD:2019jai}
Gunnar~S. Bali, Lorenzo Barca, Sara Collins, Michael Gruber, Marius L\"offler, Andreas Sch\"afer, Wolfgang S\"oldner, Philipp Wein, Simon Weish\"aupl, and Thomas Wurm.
\newblock {Nucleon axial structure from lattice QCD}.
\newblock {\em JHEP}, 05:126, 2020.

\bibitem{Park:2021ypf}
Sungwoo Park, Rajan Gupta, Boram Yoon, Santanu Mondal, Tanmoy Bhattacharya, Yong-Chull Jang, B\'alint Jo\'o, and Frank Winter.
\newblock {Precision nucleon charges and form factors using (2+1)-flavor lattice QCD}.
\newblock {\em Phys. Rev. D}, 105(5):054505, 2022.

\bibitem{Djukanovic:2022wru}
Dalibor Djukanovic, Georg von Hippel, Jonna Koponen, Harvey~B. Meyer, Konstantin Ottnad, Tobias Schulz, and Hartmut Wittig.
\newblock {Isovector axial form factor of the nucleon from lattice QCD}.
\newblock {\em Phys. Rev. D}, 106(7):074503, 2022.

\bibitem{Meyer:2022mix}
Aaron~S. Meyer, Andr\'e Walker-Loud, and Callum Wilkinson.
\newblock {Status of Lattice QCD Determination of Nucleon Form Factors and their Relevance for the Few-GeV Neutrino Program}.
\newblock {\em Ann. Rev. Nucl. Part. Sci.}, 72:205--232, 2022.

\bibitem{Jang:2023zts}
Yong-Chull Jang, Rajan Gupta, Tanmoy Bhattacharya, Boram Yoon, and Huey-Wen Lin.
\newblock {Nucleon isovector axial form factors}.
\newblock {\em Phys. Rev. D}, 109(1):014503, 2024.

\bibitem{Tomalak:2023pdi}
Oleksandr Tomalak, Rajan Gupta, and Tanmoy Bhattacharya.
\newblock {Confronting the axial-vector form factor from lattice QCD with MINERvA antineutrino-proton data}.
\newblock {\em Phys. Rev. D}, 108(7):074514, 2023.

\bibitem{Alexandrou:2023qbg}
Constantia Alexandrou, Simone Bacchio, Martha Constantinou, Jacob Finkenrath, Roberto Frezzotti, Bartosz Kostrzewa, Giannis Koutsou, Gregoris Spanoudes, and Carsten Urbach.
\newblock {Nucleon axial and pseudoscalar form factors using twisted-mass fermion ensembles at the physical point}.
\newblock {\em Phys. Rev. D}, 109(3):034503, 2024.

\bibitem{Tsuji:2023llh}
Ryutaro Tsuji, Yasumichi Aoki, Ken-Ichi Ishikawa, Yoshinobu Kuramashi, Shoichi Sasaki, Kohei Sato, Eigo Shintani, Hiromasa Watanabe, and Takeshi Yamazaki.
\newblock {Nucleon form factors in Nf=2+1 lattice QCD at the physical point: Finite lattice spacing effect on the root-mean-square radii}.
\newblock {\em Phys. Rev. D}, 109(9):094505, 2024.

\bibitem{Alexandrou:2006mc}
C.~Alexandrou, Th. Leontiou, John~W. Negele, and A.~Tsapalis.
\newblock {The Axial N to Delta transition form factors from Lattice QCD}.
\newblock {\em Phys. Rev. Lett.}, 98:052003, 2007.

\bibitem{Barca:2022uhi}
Lorenzo Barca, Gunnar Bali, and Sara Collins.
\newblock {Toward N to N\ensuremath{\pi} matrix elements from lattice QCD}.
\newblock {\em Phys. Rev. D}, 107(5):L051505, 2023.

\bibitem{Barca:2024sub}
Lorenzo Barca, Gunnar Bali, and Sara Collins.
\newblock {Progress on nucleon transition matrix elements with a lattice QCD variational analysis}.
\newblock {\em PoS}, EuroPLEx2023:002, 2024.

\bibitem{Alexandrou:2024tin}
Constantia Alexandrou, Giannis Koutsou, Yan Li, Marcus Petschlies, and Ferenc Pittler.
\newblock {Investigation of pion-nucleon contributions to nucleon matrix elements}.
\newblock {\em Phys. Rev. D}, 110(9):094514, 2024.

\bibitem{Alexandrou:2024tps}
Constantia Alexandrou, Giannis Koutsou, Yan Li, Marcus Petschlies, and Ferenc Pittler.
\newblock {Investigation of $\pi N$ contributions to nucleon matrix elements}.
\newblock {\em PoS}, LATTICE2024:317, 2025.

\bibitem{Hackl:2024whw}
Andreas Hackl and Christoph Lehner.
\newblock {Spectral analysis for nucleon-pion and nucleon-pion-pion states in both parity sectors using distillation with domain-wall fermions}.
\newblock 12 2024.

\bibitem{Blanton:2019igq}
Tyler~D. Blanton, Fernando Romero-L\'opez, and Stephen~R. Sharpe.
\newblock {Implementing the three-particle quantization condition including higher partial waves}.
\newblock {\em JHEP}, 03:106, 2019.

\bibitem{Briceno:2020vgp}
Ra\'ul~A. Brice\~no, Andrew~W. Jackura, Felipe~G. Ortega-Gama, and Keegan~H. Sherman.
\newblock {On-shell representations of two-body transition amplitudes: Single external current}.
\newblock {\em Phys. Rev. D}, 103(11):114512, 2021.

\bibitem{Liang:2019frk}
Jian Liang, Terrence Draper, Keh-Fei Liu, Alexander Rothkopf, and Yi-Bo Yang.
\newblock {Towards the nucleon hadronic tensor from lattice QCD}.
\newblock {\em Phys. Rev. D}, 101(11):114503, 2020.

\bibitem{Fukaya:2020wpp}
Hidenori Fukaya, Shoji Hashimoto, Takashi Kaneko, and Hiroshi Ohki.
\newblock {Towards fully nonperturbative computations of inelastic $\ell N$ scattering cross sections from lattice QCD}.
\newblock {\em Phys. Rev. D}, 102(11):114516, 2020.

\bibitem{Liang:2023uai}
Jian Liang, Raza~Sabbir Sufian, Bigeng Wang, Terrence Draper, Tanjib Khan, Keh-Fei Liu, and Yi-Bo Yang.
\newblock {Elastic and resonance structures of the nucleon from hadronic tensor in lattice QCD: implications for neutrino-nucleon scattering and hadron physics}.
\newblock 11 2023.

\bibitem{Hansen:2019idp}
Martin Hansen, Alessandro Lupo, and Nazario Tantalo.
\newblock {Extraction of spectral densities from lattice correlators}.
\newblock {\em Phys. Rev. D}, 99(9):094508, 2019.

\bibitem{Horak:2021syv}
Jan Horak, Jan~M. Pawlowski, Jos\'e Rodr\'\i{}guez-Quintero, Jonas Turnwald, Julian~M. Urban, Nicolas Wink, and Savvas Zafeiropoulos.
\newblock {Reconstructing QCD spectral functions with Gaussian processes}.
\newblock {\em Phys. Rev. D}, 105(3):036014, 2022.

\bibitem{Bulava:2021fre}
John Bulava, Maxwell~T. Hansen, Michael~W. Hansen, Agostino Patella, and Nazario Tantalo.
\newblock {Inclusive rates from smeared spectral densities in the two-dimensional O(3) non-linear \ensuremath{\sigma}-model}.
\newblock {\em JHEP}, 07:034, 2022.

\bibitem{DelDebbio:2024sfa}
Luigi Del~Debbio, Alessandro Lupo, Marco Panero, and Nazario Tantalo.
\newblock {Approaches to the Inverse Problem}.
\newblock In {\em {EuroPLEx Final Conference}}, 10 2024.

\bibitem{Jay:2025dzl}
William Jay.
\newblock {Approaching the Inverse Problem: Toward Lattice QCD Calculations of Inclusive Hadronic Quantities}.
\newblock In {\em {41st International Symposium on Lattice Field Theory}}, 1 2025.

\bibitem{Ji:2013dva}
Xiangdong Ji.
\newblock {Parton Physics on a Euclidean Lattice}.
\newblock {\em Phys. Rev. Lett.}, 110:262002, 2013.

\bibitem{Ma:2014jla}
Yan-Qing Ma and Jian-Wei Qiu.
\newblock {Extracting Parton Distribution Functions from Lattice QCD Calculations}.
\newblock {\em Phys. Rev. D}, 98(7):074021, 2018.

\bibitem{Radyushkin:2016hsy}
Anatoly Radyushkin.
\newblock {Nonperturbative Evolution of Parton Quasi-Distributions}.
\newblock {\em Phys. Lett. B}, 767:314--320, 2017.

\bibitem{Chambers:2017dov}
A.~J. Chambers, R.~Horsley, Y.~Nakamura, H.~Perlt, P.~E.~L. Rakow, G.~Schierholz, A.~Schiller, K.~Somfleth, R.~D. Young, and J.~M. Zanotti.
\newblock {Nucleon Structure Functions from Operator Product Expansion on the Lattice}.
\newblock {\em Phys. Rev. Lett.}, 118(24):242001, 2017.

\bibitem{Ji:2020ect}
Xiangdong Ji, Yu-Sheng Liu, Yizhuang Liu, Jian-Hui Zhang, and Yong Zhao.
\newblock {Large-momentum effective theory}.
\newblock {\em Rev. Mod. Phys.}, 93(3):035005, 2021.

\bibitem{Constantinou:2020pek}
Martha Constantinou.
\newblock {The x-dependence of hadronic parton distributions: A review on the progress of lattice QCD}.
\newblock {\em Eur. Phys. J. A}, 57(2):77, 2021.

\bibitem{Egerer:2021ymv}
Colin Egerer, Robert~G. Edwards, Christos Kallidonis, Kostas Orginos, Anatoly~V. Radyushkin, David~G. Richards, Eloy Romero, and Savvas Zafeiropoulos.
\newblock {Towards high-precision parton distributions from lattice QCD via distillation}.
\newblock {\em JHEP}, 11:148, 2021.

\bibitem{Bhattacharya:2023jsc}
Shohini Bhattacharya et~al.
\newblock {Generalized parton distributions from lattice QCD with asymmetric momentum transfer: Axial-vector case}.
\newblock {\em Phys. Rev. D}, 109(3):034508, 2024.

\bibitem{Delmar:2023agv}
Joseph Delmar, Constantia Alexandrou, Krzysztof Cichy, Martha Constantinou, and Kyriakos Hadjiyiannakou.
\newblock {Gluon PDF of the proton using twisted mass fermions}.
\newblock {\em Phys. Rev. D}, 108(9):094515, 2023.

\bibitem{Gao:2023ktu}
Xiang Gao, Andrew~D. Hanlon, Swagato Mukherjee, Peter Petreczky, Qi~Shi, Sergey Syritsyn, and Yong Zhao.
\newblock {Transversity PDFs of the proton from lattice QCD with physical quark masses}.
\newblock {\em Phys. Rev. D}, 109(5):054506, 2024.

\bibitem{Silvi:2021uya}
Giorgio Silvi et~al.
\newblock {$P$-wave nucleon-pion scattering amplitude in the $\Delta$(1232) channel from lattice QCD}.
\newblock {\em Phys. Rev. D}, 103(9):094508, 2021.

\bibitem{Pittler:2021bqw}
Ferenc Pittler, Constantia Alexandrou, Kyriakos Hadjiannakou, Giannis Koutsou, Srijit Paul, Marcus Petschlies, and Antonino Todaro.
\newblock {Elastic \ensuremath{\pi} \ensuremath{-} N scattering in the I = 3/2 channel}.
\newblock {\em PoS}, LATTICE2021:226, 2022.

\bibitem{Bulava:2022vpq}
John Bulava, Andrew~D. Hanlon, Ben H\"orz, Colin Morningstar, Amy Nicholson, Fernando Romero-L\'opez, Sarah Skinner, Pavlos Vranas, and Andr\'e Walker-Loud.
\newblock {Elastic nucleon-pion scattering at m\ensuremath{\pi} = 200 MeV from lattice QCD}.
\newblock {\em Nucl. Phys. B}, 987:116105, 2023.

\bibitem{Alexandrou:2023elk}
Constantia Alexandrou, Simone Bacchio, Giannis Koutsou, Theodoros Leontiou, Srijit Paul, Marcus Petschlies, and Ferenc Pittler.
\newblock {Elastic nucleon-pion scattering amplitudes in the \ensuremath{\Delta} channel at physical pion mass from lattice QCD}.
\newblock {\em Phys. Rev. D}, 109(3):034509, 2024.

\bibitem{Inoue:2011ai}
Takashi Inoue, Sinya Aoki, Takumi Doi, Tetsuo Hatsuda, Yoichi Ikeda, Noriyoshi Ishii, Keiko Murano, Hidekatsu Nemura, and Kanji Sasaki.
\newblock {Two-Baryon Potentials and H-Dibaryon from 3-flavor Lattice QCD Simulations}.
\newblock {\em Nucl. Phys. A}, 881:28--43, 2012.

\bibitem{Berkowitz:2015eaa}
Evan Berkowitz, Thorsten Kurth, Amy Nicholson, Balint Joo, Enrico Rinaldi, Mark Strother, Pavlos~M. Vranas, and Andre Walker-Loud.
\newblock {Two-Nucleon Higher Partial-Wave Scattering from Lattice QCD}.
\newblock {\em Phys. Lett. B}, 765:285--292, 2017.

\bibitem{Iritani:2016jie}
Takumi Iritani et~al.
\newblock {Mirage in Temporal Correlation functions for Baryon-Baryon Interactions in Lattice QCD}.
\newblock {\em JHEP}, 10:101, 2016.

\bibitem{Wagman:2017tmp}
Michael~L. Wagman, Frank Winter, Emmanuel Chang, Zohreh Davoudi, William Detmold, Kostas Orginos, Martin~J. Savage, and Phiala~E. Shanahan.
\newblock {Baryon-Baryon Interactions and Spin-Flavor Symmetry from Lattice Quantum Chromodynamics}.
\newblock {\em Phys. Rev. D}, 96(11):114510, 2017.

\bibitem{Iritani:2018vfn}
Takumi Iritani, Sinya Aoki, Takumi Doi, Tetsuo Hatsuda, Yoichi Ikeda, Takashi Inoue, Noriyoshi Ishii, Hidekatsu Nemura, and Kenji Sasaki.
\newblock {Consistency between L\"uscher\textquoteright{}s finite volume method and HAL QCD method for two-baryon systems in lattice QCD}.
\newblock {\em JHEP}, 03:007, 2019.

\bibitem{Drischler:2019xuo}
Christian Drischler, Wick Haxton, Kenneth McElvain, Emanuele Mereghetti, Amy Nicholson, Pavlos Vranas, and Andr\'e Walker-Loud.
\newblock {Towards grounding nuclear physics in QCD}.
\newblock {\em Prog. Part. Nucl. Phys.}, 121:103888, 2021.

\bibitem{Horz:2020zvv}
Ben H\"orz et~al.
\newblock {Two-nucleon S-wave interactions at the $SU(3)$ flavor-symmetric point with $m_{ud}\simeq m_s^{\rm phys}$: A first lattice QCD calculation with the stochastic Laplacian Heaviside method}.
\newblock {\em Phys. Rev. C}, 103(1):014003, 2021.

\bibitem{Amarasinghe:2021lqa}
Saman Amarasinghe, Riyadh Baghdadi, Zohreh Davoudi, William Detmold, Marc Illa, Assumpta Parreno, Andrew~V. Pochinsky, Phiala~E. Shanahan, and Michael~L. Wagman.
\newblock {Variational study of two-nucleon systems with lattice QCD}.
\newblock {\em Phys. Rev. D}, 107(9):094508, 2023.
\newblock [Erratum: Phys.Rev.D 110, 119904 (2024)].

\bibitem{Arguelles:2022tki}
C.~A. Arg\"uelles et~al.
\newblock {Snowmass white paper: beyond the standard model effects on neutrino flavor: Submitted to the proceedings of the US community study on the future of particle physics (Snowmass 2021)}.
\newblock {\em Eur. Phys. J. C}, 83(1):15, 2023.

\bibitem{Papoulias:2017qdn}
D.~K. Papoulias and T.~S. Kosmas.
\newblock {COHERENT constraints to conventional and exotic neutrino physics}.
\newblock {\em Phys. Rev. D}, 97(3):033003, 2018.

\bibitem{Coloma:2017ncl}
Pilar Coloma, M.~C. Gonzalez-Garcia, Michele Maltoni, and Thomas Schwetz.
\newblock {COHERENT Enlightenment of the Neutrino Dark Side}.
\newblock {\em Phys. Rev. D}, 96(11):115007, 2017.

\bibitem{Coloma:2023ixt}
Pilar Coloma, M.~C. Gonzalez-Garcia, Michele Maltoni, Jo\~ao~Paulo Pinheiro, and Salvador Urrea.
\newblock {Global constraints on non-standard neutrino interactions with quarks and electrons}.
\newblock {\em JHEP}, 08:032, 2023.

\bibitem{Farzan:2015doa}
Yasaman Farzan.
\newblock {A model for large non-standard interactions of neutrinos leading to the LMA-Dark solution}.
\newblock {\em Phys. Lett. B}, 748:311--315, 2015.

\bibitem{Abdullah:2022zue}
M.~Abdullah et~al.
\newblock {Coherent elastic neutrino-nucleus scattering: Terrestrial and astrophysical applications}.
\newblock {\em arXiv}, 3 2022.

\bibitem{AristizabalSierra:2024nwf}
D.~Aristizabal~Sierra, N.~Mishra, and L.~Strigari.
\newblock {Implications of first neutrino-induced nuclear recoil measurements in direct detection experiments}.
\newblock {\em arXiv}, 9 2024.

\bibitem{DeRomeri:2024iaw}
Valentina De~Romeri, Dimitrios~K. Papoulias, and Christoph~A. Ternes.
\newblock {Bounds on new neutrino interactions from the first CE$\nu$NS data at direct detection experiments}.
\newblock {\em arXiv}, 11 2024.

\bibitem{Blanco-Mas:2024ale}
Pablo Blanco-Mas, Pilar Coloma, Gonzalo Herrera, Patrick Huber, Joachim Kopp, Ian~M. Shoemaker, and Zahra Tabrizi.
\newblock {Clarity through the Neutrino Fog: Constraining New Forces in Dark Matter Detectors}.
\newblock {\em arXiv}, 11 2024.

\bibitem{Akhmedov:2008qt}
Evgeny~Kh. Akhmedov, Michele Maltoni, and Alexei~Yu. Smirnov.
\newblock {Neutrino oscillograms of the Earth: Effects of 1-2 mixing and CP-violation}.
\newblock {\em JHEP}, 06:072, 2008.

\bibitem{Arguelles:2022hrt}
C.~A. Arg\"uelles, P.~Fern\'andez, I.~Mart\'\i{}nez-Soler, and M.~Jin.
\newblock {Measuring Oscillations with a Million Atmospheric Neutrinos}.
\newblock {\em Phys. Rev. X}, 13(4):041055, 2023.

\bibitem{Akhmedov:2006hb}
Evgeny~K. Akhmedov, Michele Maltoni, and Alexei~Yu. Smirnov.
\newblock {1-3 leptonic mixing and the neutrino oscillograms of the Earth}.
\newblock {\em JHEP}, 05:077, 2007.

\bibitem{Ribordy:2013xea}
Mathieu Ribordy and Alexei~Yu Smirnov.
\newblock {Improving the neutrino mass hierarchy identification with inelasticity measurement in PINGU and ORCA}.
\newblock {\em Phys. Rev. D}, 87(11):113007, 2013.

\bibitem{Olavarrieta:2024eaq}
Santiago~Giner Olavarrieta, Miaochen Jin, Carlos~A. Arg\"uelles, Pablo Fern\'andez, and Ivan Mart\'\i{}nez-Soler.
\newblock {Boosting neutrino mass ordering sensitivity with inelasticity for atmospheric neutrino oscillation measurement}.
\newblock {\em Phys. Rev. D}, 110(5):L051101, 2024.

\bibitem{Tomalak:2024yvq}
Oleksandr Tomalak, Minerba Betancourt, Kaushik Borah, Richard~J. Hill, and Thomas Junk.
\newblock {Constraints on new physics with (anti)neutrino-nucleon scattering data}.
\newblock {\em Phys. Lett. B}, 854:138718, 2024.

\bibitem{Ilma:2024lkp}
Ilma, M.~Rafi~Alam, L.~Alvarez-Ruso, M.~Benitez Galan, I.~Ruiz~Simo, and S.~K. Singh.
\newblock {Neutrino-nucleon elastic scattering in presence of non-standard interactions: cross sections and nucleon polarizations}.
\newblock 12 2024.

\bibitem{Coloma:2017ppo}
Pilar Coloma, Pedro A.~N. Machado, Ivan Martinez-Soler, and Ian~M. Shoemaker.
\newblock {Double-Cascade Events from New Physics in Icecube}.
\newblock {\em Phys. Rev. Lett.}, 119(20):201804, 2017.

\bibitem{Atkinson:2021rnp}
Mack Atkinson, Pilar Coloma, Ivan Martinez-Soler, Noemi Rocco, and Ian~M. Shoemaker.
\newblock {Heavy Neutrino Searches through Double-Bang Events at Super-Kamiokande, DUNE, and Hyper-Kamiokande}.
\newblock {\em JHEP}, 04:174, 2022.

\bibitem{Gustafson:2022rsz}
R.~Andrew Gustafson, Ryan Plestid, and Ian~M. Shoemaker.
\newblock {Neutrino portals, terrestrial upscattering, and atmospheric neutrinos}.
\newblock {\em Phys. Rev. D}, 106(9):095037, 2022.

\bibitem{IceCube:2025kve}
R.~Abbasi et~al.
\newblock Search for heavy neutral leptons with icecube deepcore, 2025.

\bibitem{Andreopoulos:2009rq}
C.~Andreopoulos et~al.
\newblock {The GENIE Neutrino Monte Carlo Generator}.
\newblock {\em Nucl. Instrum. Meth. A}, 614:87--104, 2010.

\bibitem{Buss:2011mx}
O.~Buss, T.~Gaitanos, K.~Gallmeister, H.~van Hees, M.~Kaskulov, O.~Lalakulich, A.~B. Larionov, T.~Leitner, J.~Weil, and U.~Mosel.
\newblock {Transport-theoretical Description of Nuclear Reactions}.
\newblock {\em Phys. Rept.}, 512:1--124, 2012.

\bibitem{Juszczak:2005zs}
Cezary Juszczak, Jaroslaw~A. Nowak, and Jan~T. Sobczyk.
\newblock {Simulations from a new neutrino event generator}.
\newblock {\em Nucl. Phys. B Proc. Suppl.}, 159:211--216, 2006.

\bibitem{Golan:2012rfa}
T.~Golan, J.~T. Sobczyk, and J.~Zmuda.
\newblock {NuWro: the Wroclaw Monte Carlo Generator of Neutrino Interactions}.
\newblock {\em Nucl. Phys. B Proc. Suppl.}, 229-232:499--499, 2012.

\bibitem{Hayato:2021heg}
Yoshinari Hayato and Luke Pickering.
\newblock {The NEUT neutrino interaction simulation program library}.
\newblock {\em Eur. Phys. J. ST}, 230(24):4469--4481, 2021.

\bibitem{Isaacson:2022cwh}
Joshua Isaacson, William~I. Jay, Alessandro Lovato, Pedro A.~N. Machado, and Noemi Rocco.
\newblock {Introducing a novel event generator for electron-nucleus and neutrino-nucleus scattering}.
\newblock {\em Phys. Rev. D}, 107(3):033007, 2023.

\bibitem{Battistoni:2009jen}
G.~Battistoni, A.~Ferrari, M.~Lantz, P.~R. Sala, and G.~Smirnov.
\newblock {A neutrino-nucleon interaction generator for the FLUKA Monte Carlo code}.
\newblock In {\em {12th International Conference on Nuclear Reaction Mechanisms}}, 6 2009.

\bibitem{Gardiner:2023ejq}
S.~Gardiner, J.~Isaacson, and L.~Pickering.
\newblock {NuHepMC: A standardized event record format for neutrino event generators}.
\newblock {\em 2310.13211}, 10 2023.

\bibitem{Stowell:2016jfr}
P.~Stowell et~al.
\newblock {NUISANCE: a neutrino cross-section generator tuning and comparison framework}.
\newblock {\em JINST}, 12(01):P01016, 2017.

\bibitem{Maguire:2017ypu}
Eamonn Maguire, Lukas Heinrich, and Graeme Watt.
\newblock {HEPData: a repository for high energy physics data}.
\newblock {\em J. Phys. Conf. Ser.}, 898(10):102006, 2017.

\bibitem{Campbell:2022qmc}
J.~M. Campbell et~al.
\newblock {Event generators for high-energy physics experiments}.
\newblock {\em SciPost Phys.}, 16(5):130, 2024.

\end{thebibliography}

\end{document}